\documentclass[prd,twocolumn,nofootinbib,preprintnumbers,showpacs]{revtex4-1}

\usepackage{amsmath, mathrsfs, amssymb,amsfonts,amsthm,graphicx, epsf, dcolumn, yfonts}
\usepackage[hyperfootnotes=true]{hyperref}
\usepackage{subfigure}
\usepackage{color}
\usepackage{slashed}
\usepackage{setspace}
\pdfoutput=1
\newcommand{\be}{\begin{equation}}
\newcommand{\ee}{\end{equation}}
\newcommand{\bea}{\begin{eqnarray}}
\newcommand{\eea}{\end{eqnarray}}
\newcommand{\bwt}{\begin{widetext}}
\newcommand{\ewt}{\end{widetext}}

\begin{document}

\title{Holographic Schwinger effect in de Sitter space}
\author{Willy Fischler,$^{1,2}$ Phuc H. Nguyen,$^{2,3}$ Juan F. Pedraza$^{1,2,4}$ and Walter Tangarife$^{1,2,5}$\vspace{1mm}}

\affiliation{ \begin{spacing}{1.25}$^1$Theory Group, Department of Physics, The University of Texas, Austin, TX 78712, USA \\
$^2$Texas Cosmology Center, The University of Texas, Austin, TX 78712, USA \\
$^3$Center for Relativity, Department of Physics, The University of Texas, Austin, TX 78712, USA \\
$^4$Perimeter Institute for Theoretical Physics, Waterloo, ON N2L 2Y5, Canada \\
$^5$School of Physics and Astronomy, Tel Aviv University, Ramat-Aviv 69978, Israel\end{spacing}}

\begin{abstract}\vspace{-5mm}
Using the AdS/CFT correspondence, we construct the holographic dual of a tunneling instanton describing Schwinger pair creation in de Sitter space.
Our approach allows us to extract the critical value of the electric field for which the potential barrier disappears, rendering the vacuum unstable. In addition, we compute the large-$\lambda$, large-$N_c$ corrections to the nucleation rate and we find that it agrees with previous expectations based on perturbative computations. As a by-product of this investigation, we study the causal structure of the string dual to the nucleated pair as seen by different static observers and we show that it can be interpreted as a dynamical creation of a `gluonic' wormhole. We explain how this result provides further evidence for the ER$=$EPR conjecture as an equivalence between two descriptions of the same physical phenomenon.
\end{abstract}

\preprint{UTTG-23-14, TCC-025-14}


\maketitle

\section{Introduction}

Understanding quantum field theory in de Sitter space is of great interest in theoretical physics due to its relevance in cosmology while offering many insights about the quantum nature of spacetime \cite{Bunch:1978yq,Mottola:1984ar,bd}. An intriguing phenomenon that can be explored in curved spacetimes is the production of pairs of particles in the presence of an external electric field, the so-called Schwinger mechanism \cite{Schwinger:1951nm,Cohen:2008wz}. Even though this process is relatively well understood, some aspects of it are intriguing and deserve further investigation. In de Sitter space, in particular, the Schwinger mechanism can be employed as a framework to study false vacuum decay and is considered a focus of current research efforts \cite{Garriga:1993fh,Garriga:1994bm,Villalba:1995za,Kim:2008xv,Garriga:2012qp,Frob:2014zka,Kim:2014iba,Cai:2014qba,Kobayashi:2014zza}.

Let us briefly review what happens in flat space. In quantum electrodynamics (QED), the probability of production per volume $\mathcal{V}$ of a particle-antiparticle pair with mass $m$ and spin $j$, in a constant electric field $\vec{E}$, is
\begin{equation}
\mathcal{P}\,=\,1-{\rm exp}[\Gamma\,\mathcal{V}]\,, \label{eq:probability}
\end{equation}
where $\Gamma$ is the nucleation rate \cite{Schwinger:1951nm}:
\begin{equation}\label{Schwingerformula}
\Gamma=  \frac{(2j+1)E^2}{8\pi^3} \sum_{n=1}^\infty \frac{(-1)^{(n+1)(2j+1)}}{n^2} e^{-\frac{\pi m^2 n}{E}}\,.
\end{equation}
A qualitative understanding of this phenomenon can be obtained by looking at the potential energy of the pair in the presence of an electric field,
\begin{equation}\label{Vschwinger}
V(r) = 2m-Er-\frac{\alpha_s}{r}\,,
\end{equation}
where $\alpha_s\simeq 1/137$ is the fine-structure constant.
The electric field term generates a potential barrier and the tunneling effect creates a pair of particle-antiparticle. For small values of $E$, this effect is largely suppressed. However, as $E$ becomes larger, the barrier is lowered and the tunneling process becomes more effective. For a critical value of the electric field, $E_c$, the barrier vanishes and the vacuum becomes unstable. The critical field for the barrier above is easily found to be
\begin{equation}
E_{c} = \frac{m^{2}}{\alpha_s}\label{EcCoulomb}\,.
\end{equation}
Notice that $E_c\gg m^2$, so the critical field does not satisfy the weak-field condition, an implicit assumption for the validity of the Schwinger formula (\ref{Schwingerformula}). This suggests that the pair creation process might receive important non-perturbative contributions.

As mentioned before, the Schwinger effect has received attention given its close analogy with the process of bubble nucleation, or false vacuum decay. If we consider bubble nucleation in flat space, the bubble is momentarily at rest at the moment of nucleation and expands afterwards. However, the false vacuum has Lorentz symmetry, so it is \emph{a priori} unclear in which Lorentz frame the bubble is initially at rest. A beautiful answer was given a while ago by Coleman and De Luccia \cite{Coleman:1980aw}. They argued that the Euclidean version of the bubble must be $O(d)$ symmetric so the full solution, which arises by analytically continuing the Euclidean instanton through $t=0$, automatically respects the Minkowskian $O(d-1,1)$ invariance.
A similar situation appears for the Schwinger effect:
the Euclidean trajectory of the nucleated pair has rotational symmetry given that the electric field acts as a magnetic field in Euclidean signature and, hence, the instanton follows a usual cyclotron orbit. Moreover, the Lorentzian trajectory is automatically Lorentz invariant since it becomes a hyperbola after the analytic continuation.
This implies that the Schwinger formula (\ref{Schwingerformula}) can be recovered by considering a sum over instanton amplitudes for tunneling through the potential barrier of pair creation \cite{Affleck:1981bma}. Finally, if we consider the same problem in de Sitter space, the same machinery can be adapted in a straightforward way by replacing Lorentz invariance de Sitter invariance since $O(d,1) \supset O(d-1,1)$ \cite{Garriga:1993fh}.\footnote{We must bear in mind that in more than (1+1)-dimensions a constant electric field is not a solution of the homogeneous Maxwell's equations in de Sitter \cite{Garriga:1994bm}. The charge distribution that sources a constant electric field must be tuned and is classically unstable.}

Another motivation to study pair creation via the Schwinger mechanism is that it provides us with the perfect laboratory to explore some ideas around the recently proposed ER$=$EPR conjecture \cite{Maldacena:2013xja}. Indeed, since the nucleated pair is created from the vacuum, the pair is necessarily in a singlet state and therefore is maximally entangled.
Let us illustrate this point with a simple example. Suppose that each particle can only exist in a two-level system, with spin up $|\!\uparrow\rangle$ or spin down $|\!\downarrow\rangle$, respectively. Conservation of angular momentum dictates that the pair is always created in the singlet state:
\begin{equation}\label{singstate}
|0,0\rangle = \frac{1}{\sqrt{2}}\left(|\!\uparrow\downarrow\rangle-|\!\downarrow\uparrow\rangle\right)\,.
\end{equation}
If we denote by $S_{z1}$ and $S_{z2}$ the $z$-component of the spin of particle 1 and 2, respectively, then:
\begin{equation}
\langle S_{z1} \rangle = \langle S_{z2} \rangle = 0\,,
\end{equation}
and
\begin{equation}\label{spincorrelation}
\langle S_{z1}S_{z2} \rangle = -\frac{1}{4}\,,
\end{equation}
where the expectation values are taken in the state (\ref{singstate}). In particular, notice that
\begin{equation}
\langle S_{z1}S_{z2} \rangle \neq \langle S_{z1} \rangle \langle S_{z2} \rangle\,.
\end{equation}
The fact that the expectation value of the product $S_{z1}S_{z2}$ does not factorize into a product of expectation values is a measurement of quantum entanglement, and follows from the fact that the singlet state is not a product state.
Another quantitative measure of the quantum correlations is the position-space entanglement entropy,
or von Neumann entropy, obtained by tracing over the degrees of freedom in a region containing one of the particles.
One of the advantages that the Schwinger pair production setting offers in this context is the permanent causal disconnection: it implies that no local interaction can ever spoil the correlation (\ref{spincorrelation}).\footnote{Notice that, although permanent causal disconnection is not needed for the members of an EPR pair to be entangled, the fact that the two particles could eventually exchange signals would make the ER$=$EPR equality more subtle \cite{Chernicoff:2013iga}.} The two particles can only interact with each other by exchanging space-like photons or other quanta of the field theory. In fact, direct computation shows that summing up over all these contributions yields a value of $s=\sqrt{\lambda}$, with $\lambda\equiv g_{\mathrm{YM}}^2N_c$, for the entanglement entropy of the nucleated pair \cite{Hubeny:2014zna} (see also \cite{Lewkowycz:2013laa,Gentle:2014lva}). Thus, the fact that the nucleated pair is necessarily entangled  means that our setting is suitable to test some aspects of the ER$=$EPR equality. Some previous work along this line of research includes \cite{Jensen:2013ora,Sonner:2013mba,Chernicoff:2013iga,Gharibyan:2013aha,Seki:2014pca,Jensen:2014bpa,Jensen:2014lua}

There is an extensive literature discussing different aspects of the Schwinger effect in de Sitter space \cite{Garriga:1994bm,Villalba:1995za,Kim:2008xv,Garriga:2012qp,Frob:2014zka,Kim:2014iba,Cai:2014qba,Kobayashi:2014zza}. Naturally, most of these studies were carried out using conventional field theoretical methods, which apply to the perturbative regime of weakly coupled theories. While it is expected that such process receives important non-perturbative contributions, going beyond the weakly coupled regime is technically and conceptually challenging.
Some progress in this direction was recently initiated in \cite{Gorsky:2001up,Semenoff:2011ng}, and expanded in various directions in \cite{Ambjorn:2011wz,Bolognesi:2012gr,Sato:2013pxa,Sato:2013iua,Sato:2013dwa,Hashimoto:2013mua,Sato:2013hyw,Kawai:2013xya,Chakrabortty:2014kma}, by studying the problem in the context of the AdS/CFT correspondence \cite{Maldacena:1997re,Gubser:1998bc,Witten:1998qj}. Despite the qualitative differences between the strongly and weakly coupled regimes, certain physical quantities are remarkably similar. For instance, in the special case of $\mathcal{N}=4$ $SU(N_c)$ super-Yang-Mills (SYM) theory, in flat space, conformal symmetry dictates that the interaction potential of two charged particles has the Coulombic form (\ref{Vschwinger}), where the fine-structure constant is replaced by
\begin{equation}\label{alphacoeff}
\alpha_s = \frac{4\pi^{2}\sqrt{\lambda}}{\Gamma^{4}{(1/4)}}\,.
\end{equation}
If we trust the heuristic argument presented previously based on the effective potential, this implies that the existence of a critical field could also be derived in the strong coupling regime for SYM theory. Indeed, this expectation was confirmed in \cite{Semenoff:2011ng} by explicit construction of the dual of a tunneling instanton describing Schwinger pair creation and, thus, constitutes an actual prediction of holography. Our primary goal in this paper is to generalize these results for theories in de Sitter space.

The organization of the paper is as follows. In Section \ref{sec:setup} we describe the holographic setting we use for the description
of the Schwinger effect in de Sitter space. We emphasize the role of the UV cutoff from the bulk point of view and its significance in the dual theory.
In Section \ref{sec:potential}, we perform a potential analysis for pair creation in de Sitter space and we extract from it the critical value of the electric field. Subsequently, in Section \ref{sec:nucleation}, we construct the dual of the tunneling instanton and we estimate the strong-coupling corrections to the nucleation rate $\Gamma$. In addition, we verify the critical field obtained by the effective potential method is recovered in this framework. In Section \ref{sec:EPR}, we analyze the physical properties of the Lorentzian version of the instanton and we show that our results fit in nicely within the EPR$=$ER interpretation. In Section \ref{sec:discussion} we close with a brief summary of our findings
and conclusions.

\section{Holographic setup\label{sec:setup}}

In the present work, we focus on the description of the Schwinger effect in the strongly coupling regime, using the AdS$_{d+1}$/CFT$_d$ correspondence.
Known examples of this duality from string theory constructions
include the $d=2$, 3, 4 and 6 cases, which involve the
near-horizon geometries and low-energy worldvolume theories
of multiple D1/D5, M2, D3 and M5 branes, respectively \cite{Aharony:1999ti}.
In these setups, the bulk metric generally contains a compact manifold
that encodes the internal degrees of freedom of the dual theory.
The prototype example is the $d=4$ system, which
equates Type IIB string theory on AdS$_5\times$S$^5$ with $N_c$
units of Ramond-Ramond five-form flux through the five-sphere
to $\mathcal{N} = 4$ $SU(N_c)$ SYM theory. Replacing the sphere with other compact geometry gives rise to holographic duals of
CFTs with fewer supersymmetries. However, this compact space will play no role in our computations. For any $d$, the non-compact part of the bulk is naturally AdS$_{d+1}$, which has the same isometry group as the conformal group $SO(d,2)$.

AdS/CFT has also been used to study quantum field theory in curved space \cite{Marolf:2013ioa}.
To obtain the holographic dual of theories in de Sitter, we will use the specific construction based on the hyperbolic (or topological) AdS black hole \cite{Hutasoit:2009xy,Marolf:2010tg,Fischler:2013fba,Fischler:2014tka}. In this picture, the bulk geometry is foliated with dS slices and the metric takes the form:\footnote{More specifically, the metric (\ref{eq:metric1}) is related to the massless limit of the hyperbolic AdS black hole and is dual to the Bunch-Davies vacuum of the boundary theory. The cases with $m\neq0$ modify the falloff behavior of the bulk metric near the boundary (normalizable mode) and correspond to different states of the theory.}
\be
ds_{d+1}^2=G_{\mu\nu}dx^\mu dx^\nu=\frac{L^2}{z^2}\left[f(z)^2ds_{\text{dS}}^2+dz^2 \right]\,, \label{eq:metric1}
\ee
where
\be
f(z)\equiv \left(1-\frac{H^2z^{2}}{4}\right)\,, \label{fz}
\ee
and $ds_{\text{dS}}^2$ is the de Sitter metric in $d$ dimensions. In the above, $L$ denotes the AdS radius and $H$ the Hubble constant of the boundary theory. Note also that (\ref{eq:metric1}) is given in Fefferman-Graham form. On the other hand, the de Sitter metric can be given in any set of coordinates. We will mainly focus on the region of spacetime accessible to a single geodesic observer, \emph{i.e.} the static patch of de Sitter.
For such an observer, the metric is given by:
\be\label{staticdS}
ds_{\text{dS}}^2=-(1-H^2r^2)dt^2+\frac{dr^2}{1-H^2r^2}+r^2d\Omega^2_{d-2}\,,
\ee
and is characterized by a cosmological horizon located at $r=1/H$.
One property of the static patch is that there is a killing vector $\xi=\partial_t$ associated with the invariance under time translations, hence the name ``static''. Therefore, thermodynamic quantities such as energy, temperature and entropy are well defined, a fact that will be useful later in this work. Indeed, an observer equipped with a particle detector will detect a background of Hawking quanta at a temperature of $T_{\text{dS}}=H/2\pi$. Finally, it is worth emphasizing that the foliation used in (\ref{eq:metric1}) covers only a portion of the entire manifold, which is known as the hyperbolic patch of AdS. The Killing horizon, located at $z_H=2/H$, is analogous to a Rindler horizon, with an associated temperature and non-vanishing area. As a result, a state that is pure from the point of view of global AdS will generally be mixed
because the degrees of freedom in the hyperbolic patch will be entangled with the degrees of freedom beyond the horizon, which are traced over. This is the bulk origin of the ``thermality'' in the dual theory.\footnote{See \cite{Chowdhury:2014oba} for a discussion on the role of foliations in AdS/CFT.}

The addition of fundamental matter in the boundary theory is realized by the
introduction of a stack of $N_f$ flavor branes in the bulk geometry, whose excitations are described by open strings.\footnote{Another way to achieve this is to start with a stack of $N_c+1$ color branes and separate one from the rest. Excitations in this case transform in the fundamental of the unbroken $U(N_c)$ \cite{Semenoff:2011ng}.} We will refer to these new degrees of freedom collectively as ``quarks'', even though they generically include scalars and fermions. If $N_f\ll N_c$ we can treat these flavor fields as probes and in this limit the backreaction on the geometry can be neglected. In the boundary theory, this corresponds to work in a quenched approximation which disregards quark loops. The flavor branes will generally span all directions of the dual theory (unless we consider a defect theory), and
extend along the radial direction from the boundary at $z=0$ to a position
$z=z_m$ where they end.\footnote{This means that the part of the embedding that wraps the compact manifold degenerate to a point at $z=z_m$. This can be concretely illustrated in the D3/D7 system, where the stack of D7's wrap an $S^3\subset S^5$ that shrink to a point at $z=z_m$ \cite{Hovdebo:2005hm}.} Some remarks on the physical implications of $z_m$ are in order. First notice that $z_m\neq0$ introduces finite mass (and hence dynamical) quarks into the theory \cite{Hovdebo:2005hm}. One of the crucial consequences of having finite mass is that the quarks develop a gluonic cloud of finite Compton wavelength \cite{Chernicoff:2011vn,Agon:2014rda} which implies that the $q$-$\bar{q}$ potential would no longer be Coulombic as in (\ref{Vschwinger}). Second, notice that $z_m$ can be used as a UV regulator. According to the UV/IR connection \cite{Agon:2014rda,Susskind:1998dq,Peet:1998wn}, the bulk coordinate $z$ maps into a length scale $L\sim z$ in the boundary theory, so defining the theory at the surface $z=z_m$ is equivalent to cutting off degrees of freedom of length $L\lesssim z_m$. Finally, and perhaps most importantly, the non-normalizable modes of bulk fields (including the metric) are allowed to fluctuate at the surface $z=z_m$ and this means that we are coupling the field theory degrees of freedom to dynamical gravity. As advocated in \cite{Jensen:2014bpa}, this would be in some sense reminiscent of a Randall-Sundrum scenario.

Before proceeding further, let us mention a subtlety of the Schwinger mechanism that arises in de Sitter space. It is well known that, due to the expansion of the Universe, a constant electric field is \emph{not} a solution of the homogeneous Maxwell's equations in more than (1+1)-dimensions \cite{Garriga:1994bm} (see Appendix \ref{AppEF}). In our setup, however, such an electric field is sourced by a fundamental string density which in the probe limit can be neglected, \emph{i.e.}, it does not backreact on the background geometry. We will not be worried about the stability of the charge configuration. Instead, will continue with our analysis in arbitrary number of dimensions, bearing in mind the physical implications and possible limitations.

\section{Potential analysis for pair creation}\label{sec:potential}

As discussed in the introduction, the idea here is to perform a potential analysis of the Schwinger mechanism in de Sitter space using the tools of the AdS/CFT correspondence.
In order to do so, we need to compute the potential energy of a pair of particles (analogous to equation (\ref{Vschwinger})) in de Sitter. We will carry out the computation in two steps. In section \ref{subqqbar} we start by considering a pair of infinitely massive particles in the absence of any electric field. In this case, the holographic computation of the potential energy is a rather simple exercise but will, nevertheless, set the grounds of our computations. In section \ref{subeffqqbar} we generalize our result in two ways: first, we upgrade to the case of finite mass by cutting off the bulk geometry a distance away from the boundary and, then, we turn on a background electric field. With this result at hand, we compute the critical value of the electric field, $E_c$, and we compare our findings with the flat space result.

\subsection{Quark-antiquark potential\label{subqqbar}}

Consider a gauge theory in the static patch of de Sitter and take an infinitely heavy pair of particle-antiparticle moving along one of the orbits of the Killing vector $\xi$. We can obtain the binding energy of the pair by computing the expectation value of a rectangular Wilson loop operator \cite{Maldacena:1998im}. In gauge theories, a Wilson loop is defined as the path ordered contour integral of the gauge field,
\be\label{wloopdef}
W(\mathcal{C})=\frac{1}{N_c}\mathrm{tr}\left(\mathcal{P}e^{\oint_\mathcal{C} A}\right)\,,
\ee
where $N_c$ denotes the number of colors, the trace runs over the fundamental representation of the gauge group and $\mathcal{C}$ is a closed loop in spacetime. Intuitively, the expectation value of this operator can be thought of as the phase factor associated to the propagation of a fundamental particle around the given loop.

Let us now focus on the rectangular loop defined by $t\in[-\frac{T}{2},\frac{T}{2}]$, $x_1\equiv x\in[-\frac{\ell}{2},\frac{\ell}{2}]$ and $x_i=0$ for $i=2,...,d-2$. In the limit $T\to\infty$ the expectation value of the Wilson loop evaluates to
\be\label{expwl}
\left\langle W(\mathcal{C})\right\rangle=e^{-T E(\ell)}\,,
\ee
where $E(\ell)$ is the energy of the pair separated by a (coordinate) distance of $\ell$.\footnote{For simplicity we are considering the symmetric configuration where one of the particles is located at $x_1=-\ell/2$ and the other one at $x_1=\ell/2$. Any other configuration can be casted  as this one by transforming to the reference frame of an observer that is at equal proper length of each of the particles.} In a curved space, we expect $E(\ell)$ to have three contributions:
\be\label{threeenergy}
E(\ell)=2m+2V_{\text{grav}}(\ell/2)+V_{q\bar{q}}(\ell)\,.
\ee
where $m$ is the mass of the particles, $V_{\text{grav}}(\ell/2)$ is the gravitational potential energy of a particle at $x=\pm\ell/2$, and $V_{q\bar{q}}(\ell)$ is the binding energy of the two particles. Thus, by computing the expectation value of such a loop we can, in principle, obtain the desired potential.

\begin{figure}[t]\centering
\includegraphics[width=3.2in]{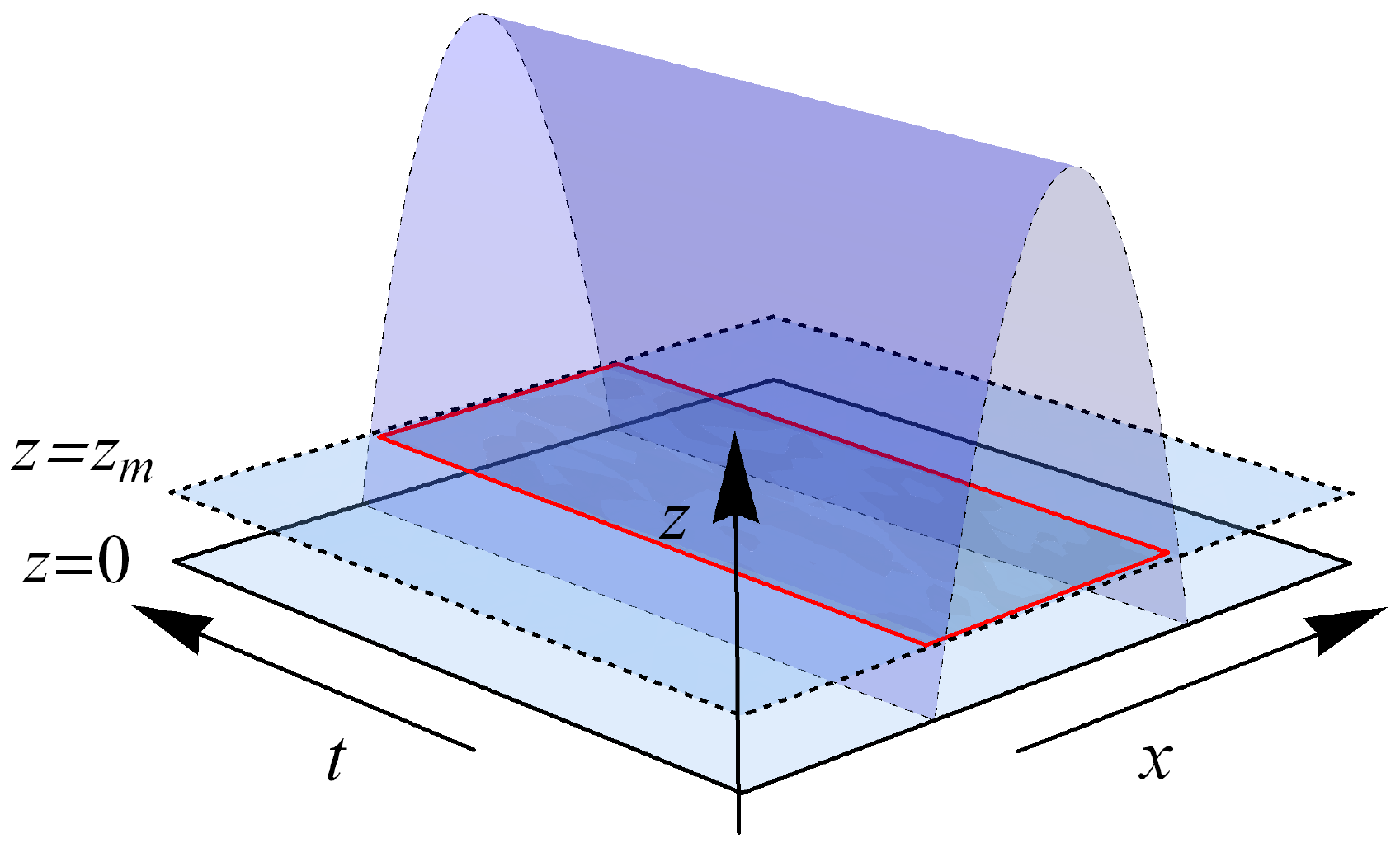}
\caption{\small Illustration of the holographic setup for the computation of the quark-antiquark potential. The red rectangle represents the Wilson loop on the probe D-brane. In the limit of infinite mass, we take the position of the brane close to the boundary, $z_m\to0$.} \label{fig:Probe3D}
\end{figure}

According to the holographic dictionary, the expectation value of a Wilson loop is given by the open string partition function,
\be\label{wloopads}
\left\langle W(\mathcal{C})\right\rangle=\int \mathcal{D}\Sigma\, e^{-S_{\text{NG}}(\Sigma)}\,,
\ee
where the integral runs over all worldsheets $\Sigma$ with boundary condition $\partial\Sigma=\mathcal{C}$ (at the position of the flavor branes, where the open strings end). Here, $S_{\text{NG}}$ is the usual Nambu-Goto action,
\be\label{ngaction}
S_{\text{NG}}\equiv\int d\tau d\sigma\mathcal{L}_{\text{NG}}=\frac{1}{2\pi\alpha'}\int d\tau d\sigma\sqrt{-\det g_{ab}}\,,
\ee
and $g_{ab}=\partial_ax^\mu(\tau,\sigma)\partial_bx^\nu(\tau,\sigma) G_{\mu\nu}$ the induced metric on the worldsheet.  In the limit of large 't Hooft coupling,\footnote{This definition is precise in AdS$_{5}$, but we will also use it for other number of dimensions.}
\be
\frac{L^2}{\alpha'}\equiv\sqrt{\lambda}\,\gg1\,,
\ee
we can make use of the saddle point approximation and (\ref{wloopads}) reduces to
\be
\left\langle W(\mathcal{C})\right\rangle=e^{-S_{\text{NG}}(\Sigma_0)}\,,
\ee
where $\Sigma_0$ is the worldsheet of minimal area subject to the boundary condition $\partial\Sigma=\mathcal{C}$ . In figure \ref{fig:Probe3D}, we plot schematically the minimal area surface for the rectangular loop we are considering.

In the static gauge, $(\tau,\sigma)=(t,x)$, the string embedding can be parameterized by a single function $z(x)$. In this case, the action (\ref{ngaction}) takes the form
\be\label{ngaction2}
S_{\text{NG}}=\frac{TL^2}{2\pi\alpha'}\int \frac{dx}{z^2}f(z)\sqrt{f(z)^2+z'(x)^2h(x)}\,,
\ee
where we defined the function
\be\label{hx}
h(x)\equiv1-H^2x^2\,,
\ee
and $f(z)$ is given in (\ref{fz}). The equation of motion that follows from (\ref{ngaction2}) is:
\bea\label{eom}
&&z''+\frac{h'(x)}{h(x)}z'+\frac{2 f(z) (f(z)-z f_z(z))}{z h(x)}\nonumber\\
&&\qquad+\frac{ 2 f(z)-3 z f_z(z)}{z f(z)}z'^2+\frac{h'(x)}{2 f(z)^2}z'^3=0\,,
\eea
where $'\equiv\partial_x$ and $f_z(x)\equiv\partial_z f(z)=-zH^2/2$. This equation is highly non-linear and, unfortunately, the general solution cannot be obtained in a closed form. For now we will proceed numerically, but later in this section we will present a parametric solution that will allow us to obtain analytic results in certain region of the parameter space.

In the numerical approach to solve equation (\ref{eom}), our goal is to find $z(x)$ subject to the boundary condition $z(\pm\ell/2)=0$.
In practice, however, it is easier to start from the IR and then implement the boundary condition through a shooting method. Due to the symmetry of the geometry, when $x=0$, $z$ reaches its maximum value $z=z_*$. Thus, we impose that
\begin{equation}
z(0)=z_*,\qquad z'(0)=0\,, \label{BCs}
\end{equation}
and then integrate numerically towards the boundary. From this solution we extract the value of $\ell$ for each $z_*$ by solving $z(\ell/2)=0$.

\begin{figure}
\includegraphics[width=8.5cm]{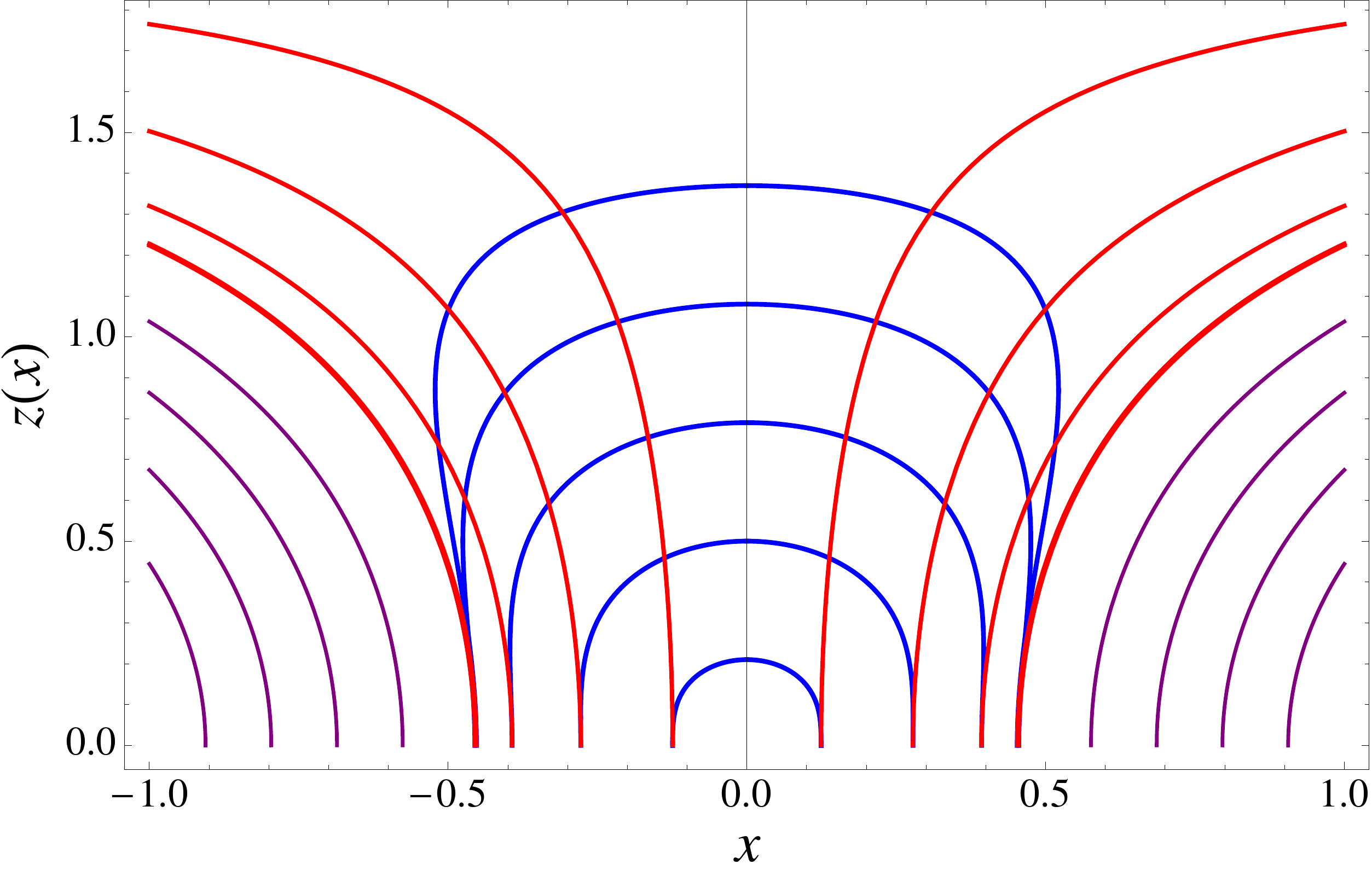}
\caption{The rectangular Wilson loops, for different separations, with $z$ and $x$ measured in units of $1/H$. The blue lines correspond to connected surfaces while the red and purple lines represent disconnected surfaces.\label{Figrectangular}}
\end{figure}

Before presenting the results, let us first discuss a subtlety of the numerical method. For large values of $z_*$ there is a point $x=x_{c}$ for which the integration breaks down before the solution reach the boundary. This is related to the fact that for such cases $z(x)$ turns out to be multivalued. Numerically, this issue can be treated as follows: first, we integrate the solution up to $x_{c}$ where $z=z_{c}$ and $z'(x_{c})=z'_{c}\to-\infty$ (this is why the numerical method breaks down). Next, we invert (\ref{eom}) in order to obtain an equation for $x(z)$. This equation is used to solve for $x(z)$ starting from $z_{c}$ and $x_{c}$ (where $x'(z_{c})=1/z'_{c}\to0$) up to the boundary. In figure \ref{Figrectangular}, we show different profiles of $z(x)$ (depicted in blue) corresponding to different values of $z_*$.

Some comments about the solutions are in order. First, for each value of $\ell$ we find that there are two solutions that satisfy the conditions (\ref{BCs}) with different values of $z_*$ (see the largest two loops in figure \ref{Figrectangular} for a particular example). We then choose the solution that yields a smaller surface area (which turns out to be the one with smaller $z_*$). Tied to this observation is the fact that $\ell$ is non-monotonic with respect to $z_*$. To be more specific, $\ell$ first increases as we increase $z_*$ from zero to critical value $z_*=z_{\text{max}}\approx1.21$ (in units of $1/H$), and then it decreases. This gives us a maximum length $H\ell_{\text{max}}\approx 0.92$  beyond which there is no solution satisfying the conditions (\ref{BCs}). Finally, for each value of $\ell$ there is also a solution to the equations of motion that corresponds to two disconnected strings with end points at $x=\pm\ell/2$. %Such a solution represents a pair of quarks that are decoupled.
In this case the profile can be obtained analytically (see Appendix \ref{AppSingleQ} for details) and the solution takes the form
\begin{equation}\label{Lorentziandisconnected}
z{(x)} = \frac{2}{H}\sqrt{\frac{x\mp\ell/2}{x\pm\ell/2}}\,.
\end{equation}
Some of these profiles are shown in figure \ref{Figrectangular} (depicted in red and purple). In cases where a connected solution exists, \emph{i.e.} for $\ell\leq\ell_{\text{max}}$, the relevant embedding is the one with minimal area.

\begin{figure}
\includegraphics[width=8.5cm]{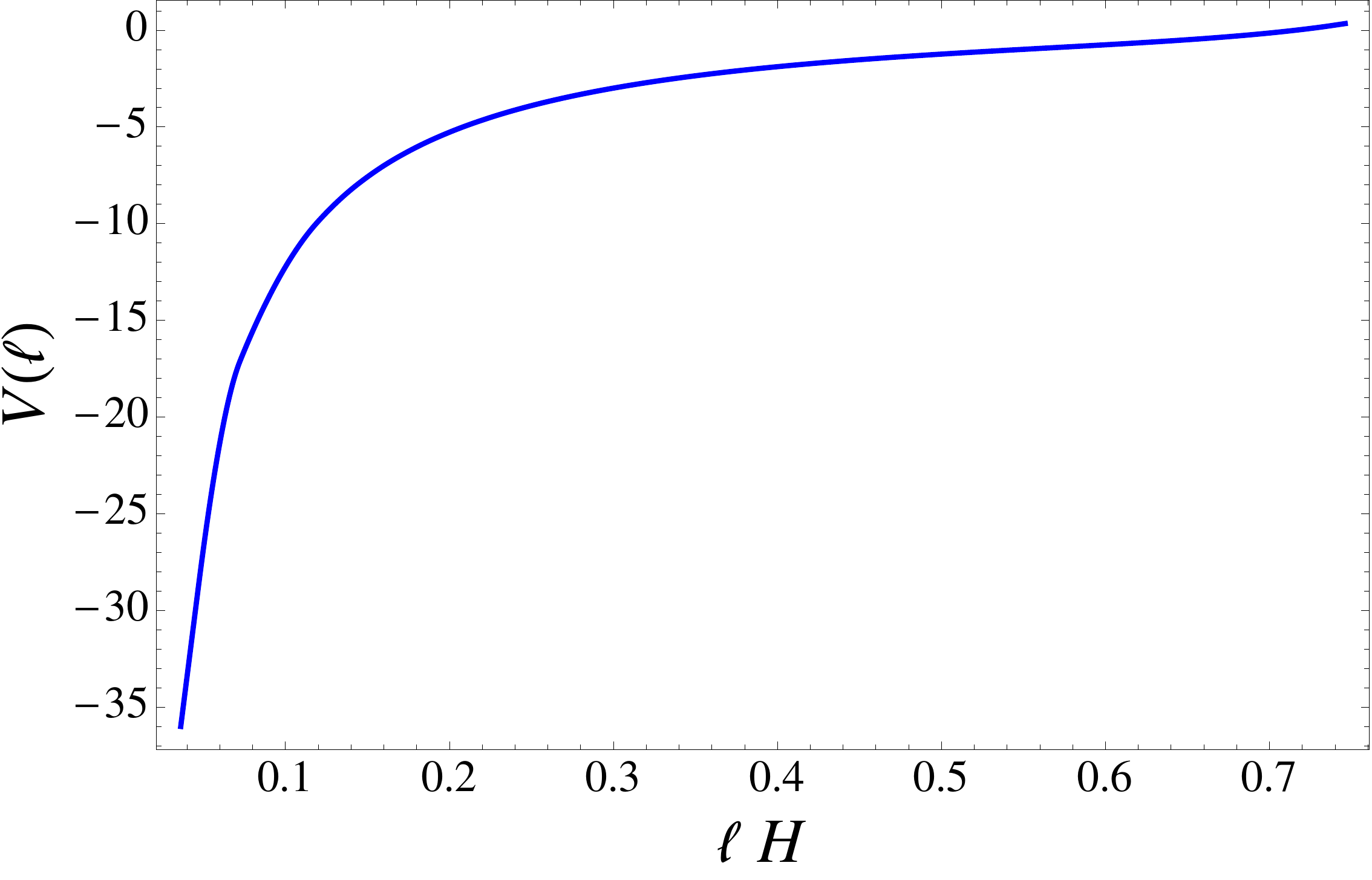}
\caption{Quark-antiquark potential obtained by evaluating the action (\ref{ngaction2}) using the numerical solution to the equation of motion (\ref{eom})\label{Potrectangular}}
\end{figure}

The numerical solution for $z(x)$ is used to evaluate the action (\ref{ngaction2}) on-shell, which gives us the energy of the pair $E(\ell)$ through (\ref{expwl}). This action is naturally divergent because it includes the intrinsic energy of the two particles, which are taken to be infinitely massive. We can easily take care of this divergence by subtracting the contribution of the disconnected solution, which includes both the mass term and the gravitational potential energy that appear in (\ref{threeenergy}) (see Appendix \ref{AppSingleQ}). The resulting potential $V_{q\bar{q}}(\ell)$ is shown in figure \ref{Potrectangular}. Similar to the finite temperature case in flat space \cite{Rey:1998bq,Brandhuber:1998bs,Brandhuber:1998er}, we find that there is a sharp transition at some $\ell=\ell_{\text{scr}}$ for which the disconnected solution becomes energetically more favorable. More specifically, for $H\ell\geq H\ell_{\text{scr}}\approx 0.75$ we observe that the potential flattens abruptly to a value of $V_{q\bar{q}}(\ell)=0$.\footnote{This ``first order'' transition is an artifact of the $N_c\to\infty$ limit and is expected to smooth out by considering $1/N_c$ corrections.} Such a transition is interpreted as the screening of the color flux tube between the two particles (which is holographically realized by the string extending between them) by the gluonic sector of the theory, in complete analogy to the phenomenon of Debye screening in classical electromagnetism.

The parametric solution for the connected worldsheet can be obtained by analytically continuing to Euclidean space (see Appendix \ref{App2} for details). The final result for the embedding can be written as
\be\label{connecsol}
Z(P) = \pm \gamma P_{0}\left[F{\left(\beta,\gamma\right)}
-\delta P_{0}^{2}\,\Pi\left(\beta,\delta,\gamma\right) \right]\,,
\ee
where $F(\beta,\gamma)$ and $\Pi(\beta,\delta,\gamma)$ are the elliptic integral of the first kind and the complete elliptic integral of the third kind, respectively, and
\bea\nonumber
\beta=\mathrm{arccos}\left(\frac{P_{0}}{P}\right)\,,\quad\gamma=\sqrt{\frac{1+P_{0}^{2}}{1+2P_{0}^{2}}}\,,\quad\delta=\frac{1}{1+P_{0}^{2}}\,.
\eea
The coordinates $(Z,P)$ are related to $(x,z)$ through
\bea
&&Z = \mathrm{arccosh}{\left(\frac{2-f(z)}{\sqrt{H^{2}z^{2}+h(x)f(z)^2}}\right)}\,,\label{statictocylinderZ}\\
&&P = \frac{f(z)}{Hz}h(x)^{1/2}\,,\label{statictocylinderP}
\eea
and $P_{0}$ is a constant of integration given by
\begin{equation}\label{P0tostaticpatch}
P_{0} = \frac{f(z_*)}{Hz_{*}}\,.
\end{equation}
Near the boundary $P\to\infty$ and $Z\to\pm Z_\infty$, where
\begin{equation}\label{Zinftostaticpatch}
Z_{\infty} = \mathrm{arccosh}{\left(\frac{1}{\sqrt{1-H^{2}l^{2}/4}}\right)}\,.
\end{equation}
By evaluating (\ref{connecsol}) at $P\to\infty$, it follows that for a given value of $\ell$ there are two minimal surfaces with different $P_0$, as expected from our numerical calculations. Among these two, the solution with larger $P_0$ is the one with minimal area (see Appendix \ref{Sectionminsurf} for explicit expressions).

In general, it is not possible to invert $P_0$ analytically as a function of $\ell$, so we cannot write down an expression for the quark-antiquark potential in a closed form. However, for small loops we can expand in $H\ell\ll1$ and perform a perturbative analysis. In this regime, we find that
\begin{equation}\label{P0smallloop}
P_{0} = \frac{A}{H\ell}\left(1-BH^{2}\ell^{2}+\mathcal{O}(H^4\ell^4)\right)\,,
\end{equation}
where
\be
A = \frac{2\sqrt{2}\pi^{3/2}}{\Gamma{\left(\frac{1}{4}\right)}^{2}}\,,\quad B = \frac{1}{12}+\frac{\Gamma{\left(\frac{1}{4}\right)}^{4}}{32\pi^{3}}+\frac{\Gamma{\left(\frac{1}{4}\right)}^{8}}{768\pi^{5}}\,.\nonumber
\ee
In Lorentzian signature, the regularized on-shell action $S_{\text{NG}}$ is related to the Euclidean action $S_{\text{NG}}^{E}$ through
\be
S_{\text{NG}}(\ell)=\frac{HT}{2\pi}S_{\text{NG}}^{E}(\ell)\,.
\ee
This follows from a simple analytic continuation.\footnote{Recall that the Euclidean time is periodic, $t_E\sim t_E+\beta$, with $\beta=2\pi/H$.}
A brief computation yields
\begin{equation}\label{qqsmall}
V_{q\bar{q}}{(\ell)} = -\frac{4\pi^{2}\sqrt{\lambda}}{\Gamma{\left(\frac{1}{4}\right)}^{4}\ell}\left(1 - CH\ell - DH^{2}\ell^2+\mathcal{O}(H^4\ell^4)\right)\,,
\end{equation}
where
\begin{equation}
C=\frac{\Gamma\left(\frac{1}{4}\right)^4}{4\pi^3}\,,\quad D = \frac{1}{12}-\frac{\Gamma\left(\frac{1}{4}\right)^{8}}{384\pi^5}\,.\nonumber
\end{equation}
%and the quark-antiquark potential is:
The leading term gives the expected Coulombic quark-antiquark potential for $\mathcal{N}=4$ SYM in flat space. In de Sitter space, however, we introduce a lengthscale $H^{-1}$ into the theory, and hence the potential receives some corrections. As a consistency check, we verified that the analytic form of (\ref{qqsmall}) matches our numerical results in the regime $H\ell\ll1$.

\subsection{Effective potential in an external electric field\label{subeffqqbar}}

Let us now discuss the finite mass case. From the bulk perspective, the inclusion of finite mass quarks amounts to imposing a radial cutoff $z_m>0$, and defining the gauge theory at this hypersurface. The value of $z_m$ is fixed by the location of the flavor branes probing the geometry and is related to the mass through (see Appendix \ref{AppSingleQ})
\begin{equation}\label{massHzm}
m = \frac{\sqrt{\lambda}}{2\pi z_{m}}\left(1-\frac{H z_m}{2}\right)^{2}\,.
\end{equation}
In the context of holography, the Schwinger effect can be understood as follows. If we turn on a background electric field $F_{tx}=E$, the DBI action for a probe brane in the geometry (\ref{eq:metric1}) is given by
\be\label{DBIaction}
S_{\rm DBI}\,=\,-T_DL^2\int d^{p+1}\xi\,\frac{f(z)^2}{z^2} \sqrt{ 1-\frac{4\pi^2z^4 E^2}{\lambda f(z)^4 }}\,.
\ee
Evaluating (\ref{DBIaction}) at $z=z_m$ we find that the action is real as long as the electric field is below the critical value:
\be\label{criticalE}
E_{c} = \frac{\sqrt{\lambda}}{2\pi z_{m}^{2}}\left(1-\frac{H^{2}z_{m}^{2}}{4}\right)^{2}\,.
\ee
For $E>E_c$, the creation of open strings is energetically favored and the system becomes unstable.
Notice that the critical field in de Sitter space is smaller than the flat space value (which can be obtained by setting $H=0$ in the equation above). This is indeed expected since the expansion of the universe is a source of particle creation, making the vacuum less stable than in flat space.

\begin{figure}[t]\centering
\includegraphics[width=3.5in]{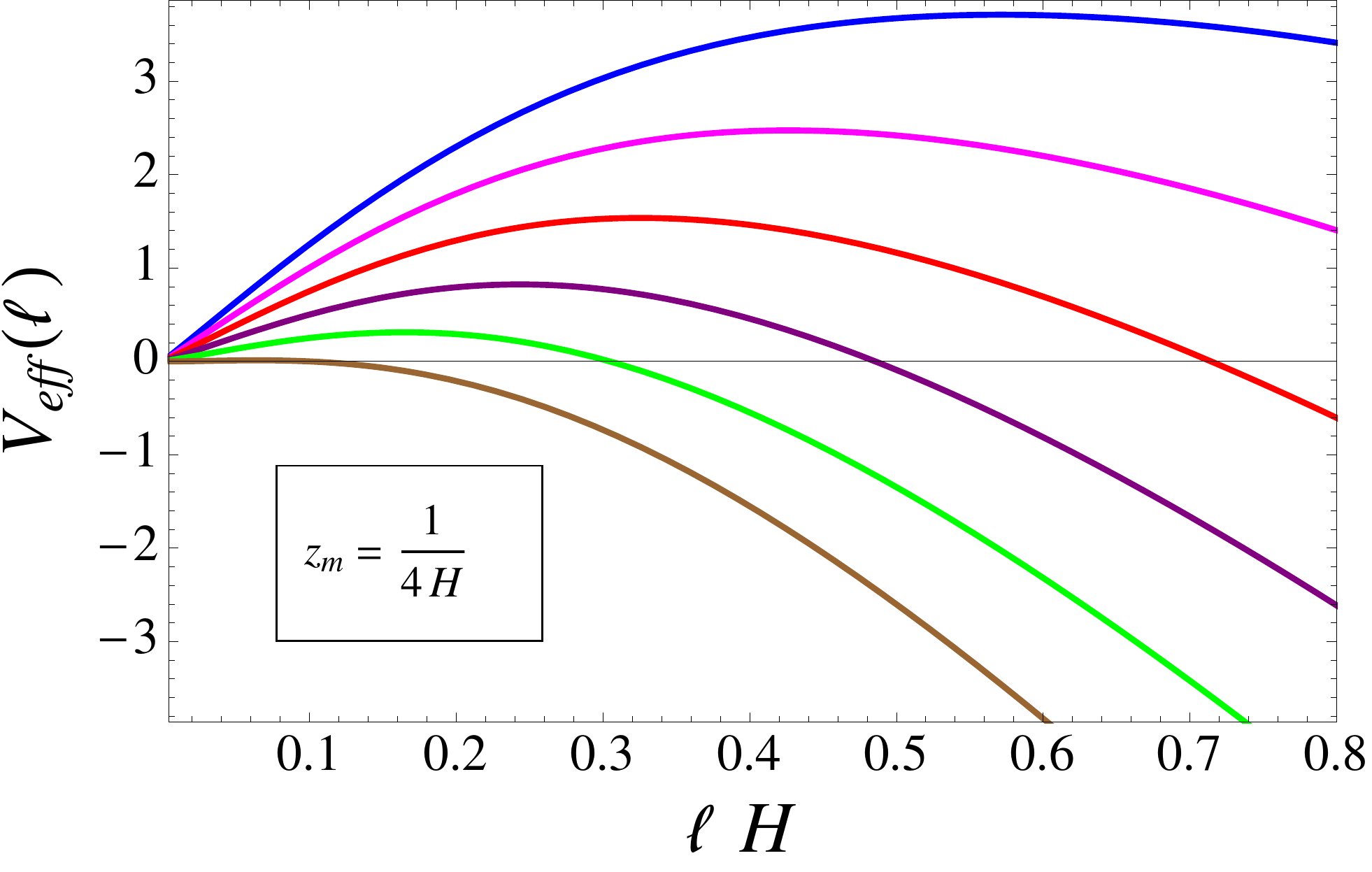}
\caption{\small Effective potential $V_{\rm eff}(x)$ for different values of $E$. At the critical value of $2\pi\alpha' \,E\approx 15.5 $ (brown line), the barrier vanishes. This agrees with the expected value from the  analysis of the DBI action.} \label{fig:Schwinger_barriers}
\end{figure}

To compute the effective potential, $V_{\text{eff}}(\ell)$, we need to recalculate the area of the connected surface but now up to the cutoff surface at $z_{m}$ (see Figure \ref{fig:Probe3D} for a schematic plot). For large enough mass $m\gg \sqrt{\lambda}H$, \emph{i.e.}, for $z_m\ll z_H$, there are still two connected surfaces for a given value of $\ell$, which is now given by
\be\label{newbczm}
z(\pm\ell/2)=z_m\,.
\ee
As before, the surface reaching deeper into the bulk has larger area and can be ignored.\footnote{For $z_m\gtrsim z_{\text{max}}\approx1.21/H$ there is only one connected surface satisfying (\ref{newbczm}). Although, it is tempting to study the physical implications of such a transition, we must keep in mind that in this regime $m\sim\mathcal{O}(\sqrt{\lambda}H)$ and, therefore, quantum corrections to the string partition function become relevant.}
Also, notice that $z_m$ can be used as a UV regulator, so there is no need to subtract the area of the disconnected surface. In this case, then, the potential $V_{q\bar{q}}^{m}(\ell)$ computed from the minimal area includes the mass of the quarks. In addition, we add the term corresponding to the contribution from the electric field $V_{E}(\ell)$ coupled to the end points of the string on the probe brane. The final result can be written as
\be
V_{\text{eff}}(\ell)=V_{q\bar{q}}^{m}(\ell)+V_E(\ell)\,.
\ee
In Figure \ref{fig:Schwinger_barriers} we show the numerical results for the effective potential corresponding to different values of $E$ and fixed mass $m$. In general, the potential barrier drops as we increase the value of $E$, as expected. For $E=E_c$ we find that $V'_{\text{eff}}(\ell)|_{\ell=0}=0$ and the barrier disappears completely.
It is noteworthy that the numerical value of $E_c$ found numerically agrees with the expected value from the DBI computation (\ref{criticalE}).
In fact, we can prove this equivalence analytically by exploring the behavior of the effective potential for small Wilson loops (following similar steps as in the previous section). This will serve as a consistency check of our numerical results. First, notice that from (\ref{statictocylinderP}) it follows that
\begin{equation}\label{pmexpp}
P_m = \frac{4-H^{2}z_{m}^{2}}{4Hz_{m}}\sqrt{1-H^{2}\ell^{2}/4}\,,
\end{equation}
where $P_m$ denotes the position of the cutoff in coordinates $(P,Z)$. Next, we expand the area of the connected surface (given in the Appendix \ref{Sectionminsurf}) around $P_m=P_0$. A brief computation yields
\be\label{intexpand}
S_{\text{NG}}^E=\frac{2 \sqrt{2} \sqrt{\lambda} P_0^{3/2} \sqrt{P_m-P_0}}{\sqrt{1+2 P_0^2}}+\mathcal{O}(P_m-P_0)\,.
\ee
Substituting the expressions for $P_0$ and $P_m$ given in (\ref{P0tostaticpatch}) and (\ref{pmexpp}) we can rewrite (\ref{intexpand}) in terms of $\ell$, $z_m$ and $z_*$. We can get rid of the $z_*$ dependence by expanding the solution for the embedding (\ref{connecsol}) around $P=P_0$ and evaluating at $P_m$. In terms of the original variables, we get
\be\label{zstzmell}
z_*=z_m-\frac{\ell^2}{4z_m}\left(1-\frac{H^4 z_m^4}{16}\right)^2+\mathcal{O}\left(\frac{\ell^4}{z_m^2}\right)\,.
\ee
Plugging (\ref{zstzmell}) into (\ref{intexpand}) we obtain
\begin{equation}
S_{\text{NG}}^E = \frac{\sqrt{\lambda}\ell}{Hz_{m}^{2}}\left(1-\frac{H^{2}z_{m}^{2}}{4}\right)^{2}\left[1 + \mathcal{O}\left(\frac{\ell}{z_m}\right)\right]\,.
\end{equation}
Finally, analytically continuing to Lorentzian signature and adding the contribution from the electric field, $V_E\simeq-E\ell$ , we arrive to:
\begin{equation}\label{eq:veffEc}
V_{\text{eff}}{(\ell)} \simeq \left(E_c-E\right)\ell + \cdots\,,
\end{equation}
where $E_c$ is given by (\ref{criticalE}). This result matches our previous expectation and serves as a non-trivial check of our numerical results.

\section{The Euclidean instanton and the nucleation rate}\label{sec:nucleation}

The nucleation rate of a particle-antiparticle pair (\ref{Schwingerformula}) was originally obtained by considering the contribution of the appropriate Feynman diagrams \cite{Schwinger:1951nm}. Later, it was realized that the same could alternatively be derived from the imaginary part of the Euclidean world-line path integral \cite{Affleck:1981bma}. For $j=0$, this method amounts to compute
\be\label{eucnrate}
\Gamma=-\frac{2}{\mathcal{V}}\Im\int \frac{dT}{T} \int \mathcal{D}x^\mu e^{-S_E[T,x^\mu]}\,,
\ee
where
\be\label{actweak}
S_E[T,x^\mu]=\int_0^1d\tau\left(\frac{\dot{x}^2}{4T}+m^2T-i A_{\mu}\dot{x}^\mu \right)\,,
\ee
and $T$ is a Lagrange multiplier. Recall that in Euclidean signature the time coordinate is periodic so in (\ref{eucnrate}) it is understood that the integral runs over world-lines satisfying periodic boundary conditions $x^{\mu}(\tau+1)=x^{\mu}(\tau)$.

The fact that the Schwinger formula (\ref{Schwingerformula}) is a sum of exponentials suggests that the production rate can be obtained in Euclidean signature as a sum over instanton amplitudes for tunneling through the potential barrier of pair creation. The explicit computation was carried out in \cite{Affleck:1981bma}, but it is easy to see that it indeed yields the correct answer for the exponential factor in the Schwinger formula; in Euclidean space, the electric field acts as a magnetic field and an instanton of the world-line path integral is a cyclotron orbit that wraps $n$ times the time direction. A brief calculation shows that the contribution of each of these is given by
\be
S_n=\frac{\pi m^2 n}{E}\,,
\ee
which correctly reproduces the factor in (\ref{Schwingerformula}). The rest of computation is technically more involved, as it requires to perform the full path integral over fluctuations of the classical solutions. Needless to say, the final result agrees with the Schwinger formula, including the prefactor of the exponential.

In the large-$N_c$ and large-$\lambda$ regime, we expect (\ref{actweak}) to be modified due to several factors. For example, interaction with the adjoint degrees of freedom of the gauge theory, which are usually ignorable at weak coupling, must be now taken into consideration. These corrections can be accounted for by the inclusion of a Wilson loop amplitude $W(x)$ in the path integral \cite{Semenoff:2011ng},
\be\label{actweak2}
S[T,x^\mu]\to S[T,x^\mu]-\log W(x^\mu)\,,
\ee
where $W(x^\mu)$ is defined as in (\ref{wloopdef}).\footnote{Depending on the specific theory, the Wilson loop operator has to be modified (\emph{e.g.} including couplings to other fields) in order to preserve gauge invariance.}
Our goal is to compute the contribution of the Wilson loop using the tools of the gauge/gravity correspondence \cite{Maldacena:1998im}.
Here, we will only deal with the exponential factor of the nucleation rate $\Gamma\sim e^{-S_E}$ for which a classical bulk computation is valid, but we leave the study of quantum fluctuations for future studies. We will compare with the weak coupling results (see \emph{e.g.} \cite{Garriga:1993fh}) which we revisit in Appendix \ref{App}.

\begin{figure}[t!]\centering
\includegraphics[width=2.5in]{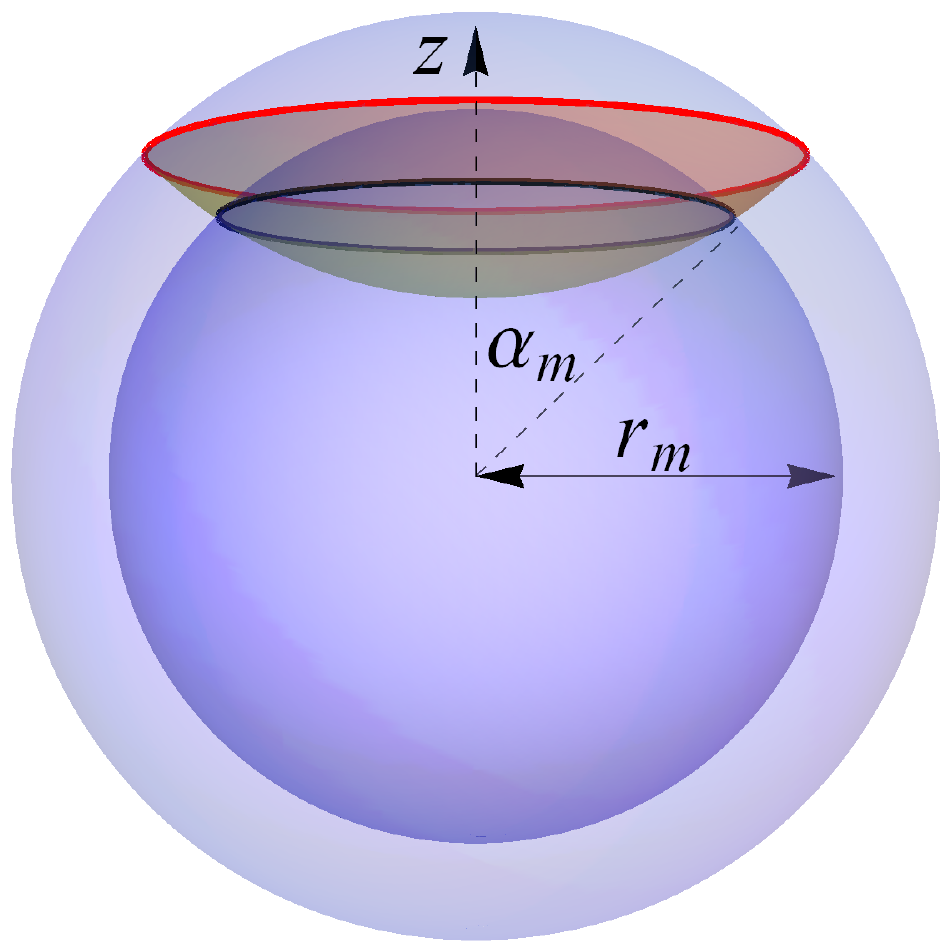}
\caption{Holographic dual of the Euclidean instanton describing Schwinger pair creation in de Sitter space. The outer sphere is the boundary of the Poincar\'{e} ball. The inner sphere denotes the location of the probe brane, which is placed at $r = r_{m}$. The circular loop describing the instanton is at $\theta = \alpha_m$ (depicted in black), and the minimal surface is the inner portion of a spherical cap that intersects the loop and reaches the boundary at a right angle.}\label{fig:Fig11}
\end{figure}

We will parametrize the Euclidean AdS in coordinates of the Poincar\'e ball, $(r,\theta,\phi)$, where $r$ denotes the bulk radial direction and $\theta$ and $\phi$ are arbitrary polar and azimuthal angles (see Appendix \ref{App2} for details). As usual, we consider a Wilson loop ending on some probe brane at $r=r_{m}$ rather than on the boundary in order to avoid infinite mass. According to (\ref{massHzm}) and (\ref{eqnB2}), this cutoff is related to the mass through
\be\label{rmzmrel}
r_m = \frac{2-Hz_m}{2+Hz_m}= \left(1+\frac{\sqrt{\lambda}H}{m\pi}\right)^{-1/2}\,.
\ee
In this case, the relevant surface representing the Euclidean instanton is a spherical cap that intersects the boundary of the ball at right angle (\ref{discsolPB}). In Figure \ref{fig:Fig11} we show schematically the setup for this configuration.
The Nambu-Goto action is proportional to the area of this surface truncated at $r_{m}$, and can be easily obtained from (\ref{areadisconnected}). The result is:
\be
S_{\text{NG}}^E = n\sqrt{\lambda}\left(\sqrt{1 + \frac{4 r_{m}^{2}}{(1-r_{m}^{2})^{2}}\sin^{2}{\alpha_{m}}}-1\right)\,,
\ee
where $n$ is the winding number and $\alpha_m$ is the polar angle subtended by the loop from the axis of symmetry. This angle $\alpha_m$ is related to the angle $\alpha$ at the boundary through
\be\label{alphamdef}
\cos\alpha_m = \frac{1+r_m^2}{2 r_m}\cos\alpha=\left(\frac{4+H^{2}z_{m}^{2}}{4-H^{2}z_{m}^{2}}\right)\cos{\alpha}\,.
\ee
Following \cite{Semenoff:2011ng}, we add to the action the interaction with the EM field,
\be
S_E=S_{\text{NG}}^E+S_{\text{EM}}^E \,,
\ee
where
\begin{equation}
S_{\text{EM}}^E = -\frac{2\pi En}{H^{2}}(1-\cos{\alpha}_m)\,.
\end{equation}
The derivation of this expression is presented in Appendix \ref{App}. Extremizing the total action with respect to $\alpha_m$ yields
\be\label{extremalsintheta}
\sin^2{\alpha}_m =\frac{4\lambda  H^4 r_m^4-E^2 \pi ^2 \left(1-r_m^2\right)^4}{4 \lambda  H^4 r_m^4+4 E^2 \pi ^2
r_m\left(1-r_m^2\right)^2}\,,
\ee
which implies the existence of a critical electric field,
\begin{equation}
E_{c} = \frac{2\sqrt{\lambda}H^{2}r_{m}^{2}}{\pi(1-r_{m}^{2})^{2}}\,.
\end{equation}
In terms of $z_{m}$, this expression is equivalent to the critical field found from the DBI result (\ref{criticalE}) and thus, serves as an consistency check of our computations. We can now rewrite the most probable separation between the quark and anti-quark and the Nambu-Goto action in terms of the critical field and boundary parameters:
\begin{equation}\label{mostprobablesintheta}
\sin^2{\alpha_m} = \frac{ \sqrt{\lambda}H^2(E_c^2-E^2) }{E_c(\sqrt{\lambda }H^2E_c +2\pi E^2)}\,,
\end{equation}
and
\be\label{NGCircularWloop}
S_{\text{NG}}^E =n\sqrt{\lambda}\left(\sqrt{1 + \frac{4 m^{2}\pi^{2}}{\lambda H^{2}}\left(1+\frac{\sqrt{\lambda}H}{m\pi}\right)\sin^{2}{\alpha_{m}}}-1\right).
\ee
Notice that as $E\to E_{c}$, both $\alpha_m$ and $S_{\text{NG}}^E$ approach to zero, indicating the disappearance of the tunneling barrier. Also, in the same limit, the total action $S_E$ vanishes and the summation over $n$ is unsuppressed.

It is instructive to consider the following two limits. First, consider the flat space limit: in this case, we have to convert to the variable $y = H^{-1}\sin{\alpha_m}$, and then take the limit $H \rightarrow 0$. The result is:
\begin{equation}
S_{\text{NG}}^{E(\text{flat})} = n\sqrt{\lambda + 4m^{2}\pi^{2}y^{2}} -n\sqrt{\lambda}\,,
\end{equation}
in agreement with \cite{Semenoff:2011ng}. Second, consider the case $H\neq0$ and $m \gg \sqrt{\lambda}H$.
In this regime, we obtain
\begin{equation}\label{NGSE}
S_{\text{NG}}^{E} \simeq \frac{2\pi mn}{H}\sin{\alpha_m} - n\sqrt{\lambda} (1-\sin{\alpha_m})+\cdots\,,
\end{equation}
where the dots are terms suppressed by extra factors of $\sqrt{\lambda}H/m$.
Notice that the first term above correctly reproduces the weak coupling result (\ref{Euclideanaction}). The second term, on the other hand, corresponds to contribution coming from the Wilson loop in (\ref{actweak2}) and can be thought of as a large-$N_c$, large-$\lambda$ correction to the action.

Finally, in order to obtain the production rate we substitute the extremal value of $\sin{\alpha_m}$ into the total action $S_E$. The general result is a cumbersome expression, but we will explicitly write down the result in the limit $m \gg H\sqrt{\lambda}$, for which our classical calculation is valid:
\begin{equation}\label{productionrate}
S_E = \left( \frac{2\pi n m}{H}-\frac{2\pi nE}{H^{2}} \right) \left(1+ \frac{E^{2}}{2H^{2}m^2} + \cdots \right).
\end{equation}
If we turn off the large-$N_c$ corrections, then we can recognize in this expression two contributions to the nucleation rate $\Gamma\sim e^{-S_{E}}$. The first term is the usual Boltzmann factor with the de Sitter temperature, $T_{\text{dS}}=H/2\pi$, and therefore is the production rate due to the expansion of space. The second term is due to the proper Schwinger effect and is linear in $E$. Finally, if we compare with the equivalent at weak coupling (\ref{productionratewk}), we find exact agreement in the regime $m \gg H$. This implies that the terms in (\ref{productionrate}) containing higher powers in the electric field (which come from the Wilson loop contribution) can be interpreted as the non-perturbative correction to the nucleation rate. We also point out that our result agrees with the semi-classical analysis presented in \cite{Frob:2014zka}. The production rate computed in that paper is
\begin{equation}\label{FrobNR}
S_E = 2\pi n \left(\frac{1}{H}\sqrt{m^{2}-\frac{H^{2}}{4}} - \frac{E}{H^{2}} \right)\,,
\end{equation}
which reduces to (\ref{productionrate}) in the regime $m \gg H$.

\section{Remarks on the ER$=$EPR conjecture}\label{sec:EPR}

In a recent paper \cite{Maldacena:2013xja}, Maldacena and
Susskind made the observation that configurations of black holes connected by
a (non-traversable) wormhole, or Einstein-Rosen bridge, should be interpreted as states where
the black holes are maximally entangled. They conjectured that this relation might hold in more general entangled systems
so that even a single EPR pair would be connected by a Planck-scale wormhole encoding their entanglement. Jensen and Karch \cite{Jensen:2013ora} gave further evidence in support of this conjecture by
taking the EPR pair to be a color-singlet quark-antiquark pair (in $\mathcal{N}=4$ SYM) undergoing constant acceleration and showing that its holographic dual has a wormhole.

Later in \cite{Sonner:2013mba}, it was argued that the configuration studied in \cite{Jensen:2013ora} is nothing but the Lorentzian continuation of the instanton associated with Schwinger pair creation, thus making contact with the
black hole pair production scenario discussed in \cite{Maldacena:2013xja}. Accordingly,
in this section we will study the Lorentzian worldsheet associated to the nucleated quark-antiquark pair in de Sitter space. We argue that the fact that its causal structure resembles a two-sided black hole connected by a (non-traversable) wormhole
provides supporting evidence in favor of  the ER$=$EPR conjecture.

In global coordinates, the Lorentzian worldsheet can be obtained by analytically continuing the solution we used in the previous section (\ref{discsolPB}) (see Figure \ref{fig:Fig11}).
We will define a coordinate $w = \frac{1}{H}\sin{\phi}$ so that
\be
ds_{\text{dS}}^2=-d\tau^{2}+\frac{\cosh^{2}{(H\tau)}}{1-H^{2}w^{2}}dw^{2}\,.
\ee
Moreover, we will transform back to the Fefferman-Graham coordinate $z$ defined in (\ref{eq:metric1}). The advantage of the coordinate $w$ over the angle $\phi$ is that we can easily take the limit $H\to0$ and recover the Poincar\'{e} patch of AdS foliated with Minkowski slices.

In the coordinate system described above, the worldsheet embedding takes the following form:
\begin{equation}\label{Lorentzianworldsheet}
z=\frac{2\sqrt{1-H^{2}w^{2}}\cosh{(H\tau)}-2\sqrt{1-H^{2}w_{0}^{2}}}{H\sqrt{(1-H^{2}w^{2})\cosh^{2}{(H\tau)}-(1-H^{2}w_{0}^{2})}}\,,
\end{equation}
where $w_{0} = \frac{1}{H}\sin{\alpha}$. As a consistency check, notice that taking the flat space limit yields
\begin{equation}\label{flatspacews}
z = \sqrt{\tau^{2}-w^{2}+w_{0}^{2}}\,.
\end{equation}
This is exactly the string profile for an accelerated quark-antiquark pair in Minkowski space found by Xiao \cite{Xiao:2008nr} and considered by Jensen and Karch in \cite{Jensen:2013ora}. The worldline  of the nucleated pair $w_{q}{(\tau)}$ can be found by setting $z=z_m$ in (\ref{Lorentzianworldsheet}):\footnote{Notice that our solutions differ from the ones studied by Jensen, Karch and Robinson in \cite{Jensen:2014bpa}. In their case, the worldlines of the quark-antiquark pair follow contant-$\phi$ trajectories.}
\begin{equation}\label{quarktrajectory}
w_{q}{(\tau)} = \pm \frac{1}{H} \sqrt{1-\cos^{2}{\alpha_{m}}\mathrm{sech}^{2}{(H\tau)}}\,,
\end{equation}
where $\alpha_m$ is given in (\ref{alphamdef}). By evaluating (\ref{quarktrajectory}) at $\tau=0$, we can see that the parameter $\alpha_m$ (which indicates the size of the circle in the Euclidean signature at $r=r_m$) sets the initial separation in $\phi$ between the two particles.
At late times, on the other hand, $w_q \to \pm H^{-1}$. This means that the two particles approach opposite sides of the de Sitter hyperboloid.
Finally, we point out the fact that the worldsheet never penetrates deeper than $Hz = 2$, \emph{i.e.} it stays within the hyperbolic patch of AdS.

As pointed out by Jensen and Karch in \cite{Jensen:2013ora}, the flat-space worldsheet (\ref{flatspacews}) has a two-sided horizon at $z^{ws}_{H}=w_{0}$ and is therefore a wormhole. We will now show that worldsheet (\ref{Lorentzianworldsheet}) also possesses the causal structure of a wormhole for $H\neq0$. Taking $(\phi,\tau)$ as the coordinates on the worldsheet, the induced metric has the following components:
\be
\begin{split}
&h_{\tau\tau} = \frac{L^{2}}{z^{2}}\left(\dot{z}^{2}-f{(z)}^{2}\right)\,,\qquad h_{\tau\phi} = \frac{L^{2}}{z^{2}}z'\dot{z}\,,\\
&\quad\;\, h_{\phi\phi} = \frac{L^{2}}{z^{2}}\left(z'^{2}+\frac{f(z)^{2}}{H^{2}}\cosh^{2}{(H\tau)}\right)\,,
\end{split}
\ee
where $\dot{\,}\equiv\partial_\tau$ and $'\equiv\partial_\phi$. The geodesic equation for a null geodesic in this geometry reads
\begin{equation}
h_{\tau\tau}  + 2h_{\tau\phi}\left(\frac{d\phi}{d\tau}\right)+ h_{\phi\phi}\left(\frac{d\phi}{d\tau}\right)^{2}=0\,.
\end{equation}
We will solve this equation numerically. If the geodesic intersects the worldline of the quark at time $\tau_{0}$, then we have the boundary condition
\begin{equation}
\phi{(\tau_{0})} = \arccos{\left[\left(\frac{4-H^{2}z_{m}^{2}}{4+H^{2}z_{m}^{2}}\right)\cos{\alpha_{m}}\,\mathrm{sech}{(H\tau_{0})}\right]}\,.
\end{equation}
\begin{figure}[t!]
\centering
\includegraphics[width=6cm]{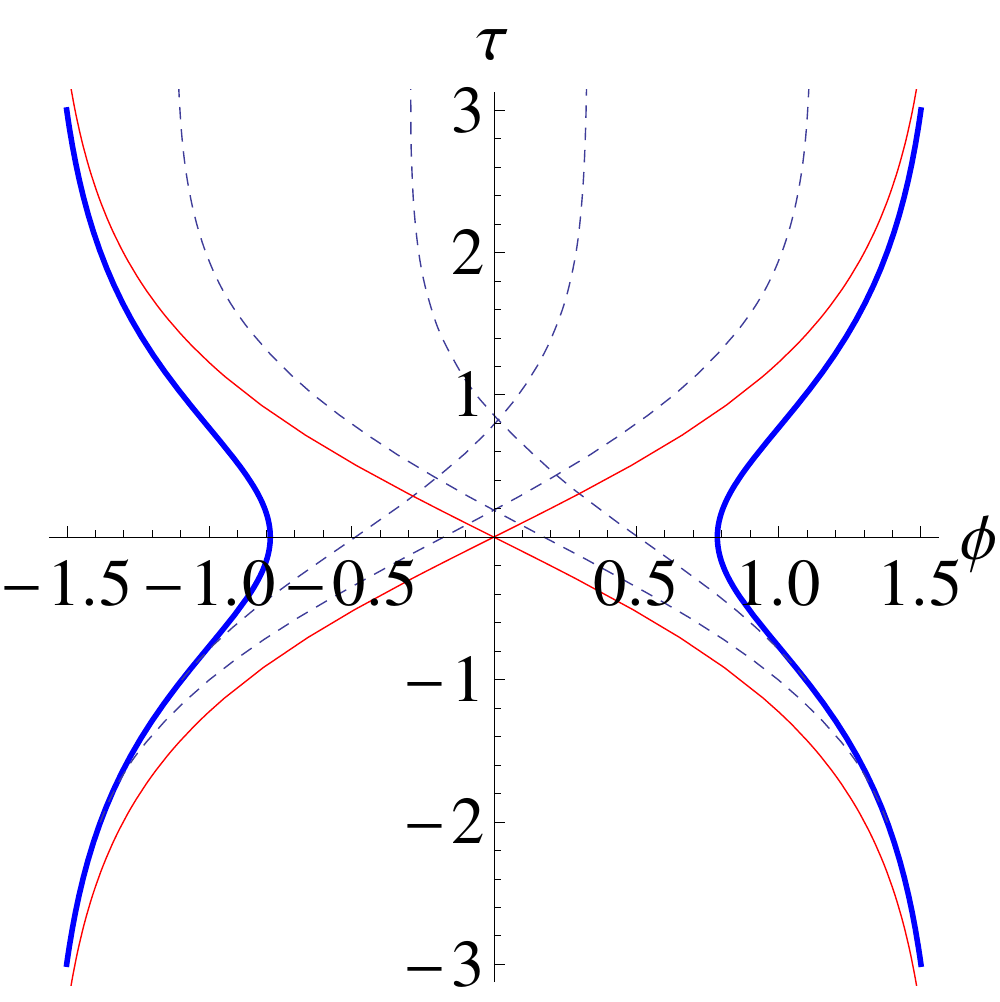}
\caption{The causal structure of the worldsheet geometry. The vertical axis is the global time $\tau$, and the horizontal axis is the angle $\phi$. The thick blue lines correspond to the worldlines of the nucleated pair. The dashed blue lines are outgoing null geodesics emitted from the two boundaries. The red lines represent the worldsheet horizons. Both $\tau$ and $\phi$ are measured in units of $H$ which we set to unity. For this example we have chosen $\alpha_m=\pi/4$ and $z_m=1/10$. \label{conformalstructure}}
\end{figure}

The two edges of the worldsheet play a role analogous to the boundary of an asymptotically AdS space: the null geodesics can reach the edges and bounce back to the interior of the worldsheet (with appropriate boundary conditions). However, a single null geodesic cannot connect the two AdS boundaries at finite $\tau$. This can be seen from Figure \ref{conformalstructure}, in which we plot the null geodesics emitted from the two boundaries for different choices of $\tau_{0}$. By inspection, we can see that the worldsheet is divided into four causally distinct regions, separated one from another by a worldsheet horizon (depicted in red). This means that the conformal structure of the worldsheet is exactly the same as that of a 2-sided black hole in AdS, as we previously advertised.

We can obtain an explicit expression for the worldsheet horizon: it consists of the null geodesics passing through $\phi=0$ at $\tau=0$, and reaching $\phi = \pm \frac{\pi}{2}$ at $\tau \rightarrow \pm\infty$. After some algebra, we find that the horizon lies at a fixed radial depth:
\be\label{wshorizon}
\begin{split}
 z^{ws}_{H} &= \frac{2}{H}\sqrt{\frac{4+H^{2}z_{m}^{2}-(4-H^{2}z_{m}^{2})\cos{\alpha_{m}}}{4+H^{2}z_{m}^{2}+(4-H^{2}z_{m}^{2})\cos{\alpha_{m}}}}\,,\\
&=\frac{2}{H}\left(\frac{1-\cos{\alpha}}{\sin{\alpha}}\right)\,.
\end{split}
\ee
A few comments are in order. First, notice that in the flat-space limit we recover the known result $z^{ws}_{H}\to w_{0}$, as expected. More importantly, if we take $z_m\geq z^{ws}_{H}$ one would naively think that the wormhole disappears. If this were true, it would be in contradiction with the ER$=$EPR interpretation given that, although the EPR pair would still be entangled (regardless the value of $z_m$), the wormhole would no longer be present. However, notice that $z_m\geq z^{ws}_{H}$ is not allowed! In such case $\alpha_m$ would be imaginary regardless the value of $\alpha$ and the trajectory of the quark-antiquark pair (\ref{quarktrajectory}) would be space-like.\footnote{This does not imply that there is a bound on the quark mass for a given acceleration (or electric field). In other words, if we first fix $z_m$, then we can always find an $\alpha$ such that $\alpha_m\in(0,\pi/2)$.} Of course, such an embedding would not be physically relevant in the context we are considering. On the other hand, notice that even for $z_m<z^{ws}_{H}$ the portion of the string above the horizon moves locally faster than the speed of light. This should not be surprising since from the gravity point of view the coordinates $(\tau,\phi)$ become space-like/time-like in that region of the worldsheet. From the field theory perspective, this is related
to the fact that the quark and the antiquark are causally disconnected, so (at least part of) the flux tube that connects them must stretch faster than the speed of light. In this sense, the wormhole subtended by the worldsheet of the EPR pair is interpreted as a `gluonic' wormhole.\footnote{We thank Alberto G\"uijosa for a discussion on this point.}

\subsection{The viewpoint of static observers\label{substatic}}

\begin{figure}[t!]\centering
\includegraphics[width=3.4in]{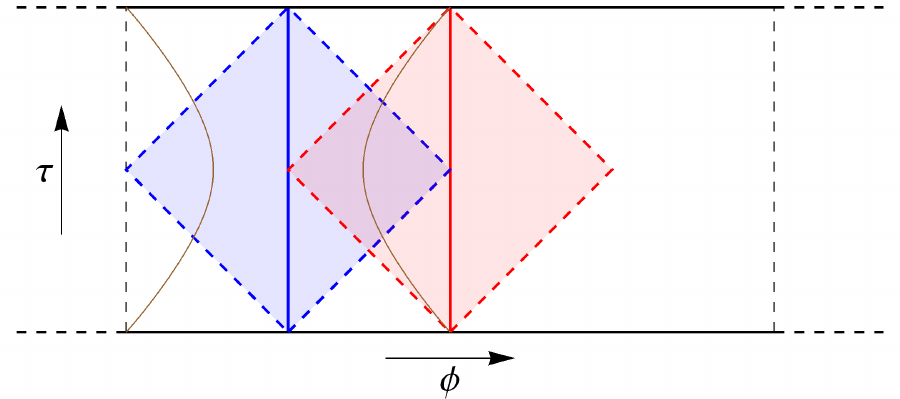}
\caption{Penrose diagram of de Sitter space. The global coordinates $(\tau,\phi)$ cover the whole manifold, with $\tau$
being the vertical axis and $\phi$ the horizontal. The coordinate $\phi$ has period of $2\pi$, so the two vertical dashed lines must be periodically identified.
The brown lines represent the worldlines of the quark-antiquark pair. The blue and red vertical lines correspond to the worldlines of the EPR observer and the Hawking observer, respectively, and the shaded regions are the causal diamonds of these observers. From the point of view of the EPR observer, the two particles enter the diamond at static time $t\to-\infty$ and then exit at $t\to\infty$. The Hawking observer only has access to one of the particles while the other always lies behind its horizon.
}\label{fig:penrose}
\end{figure}

Since no observer in de Sitter space has access to the entire manifold, it is instructive to transform the worldsheet into static coordinates adapted to different geodesic observers and study how their causal structure affect the observations of the quark-antiquark pair. First, we will consider an observer at a fixed $\phi$, equidistant from the two particles. From the point of view of this observer the two particles are interpreted as an usual EPR pair nucleated from the Bunch-Davies vacuum of de Sitter. Importantly, in global coordinates the two particles actually stay a finite time $\tau$ inside the causal diamond of this ``EPR observer''. The second observer we will consider is also at a fixed $\phi$ but is in causal contact with only one of the two particles, while the other always lies behind its horizon. We will refer to the second observer as the ``Hawking observer'' since, with respect to this observer, the particles can be interpreted as a Hawking pair nucleated from the cosmological horizon. The worldlines of these observers and their associated causal diamonds are better visualized with the help of a Penrose diagram, as shown in Figure \ref{fig:penrose}.

\begin{figure*}
\centering
\hskip -0.12in
\includegraphics[width=8cm]{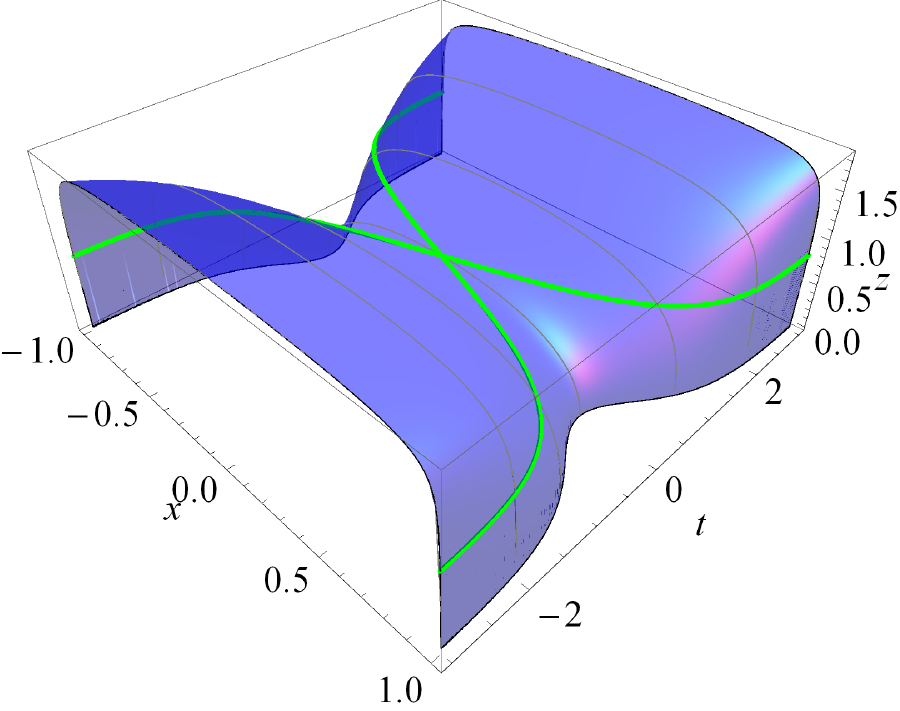}
\hskip 0.6in
\includegraphics[width=8cm]{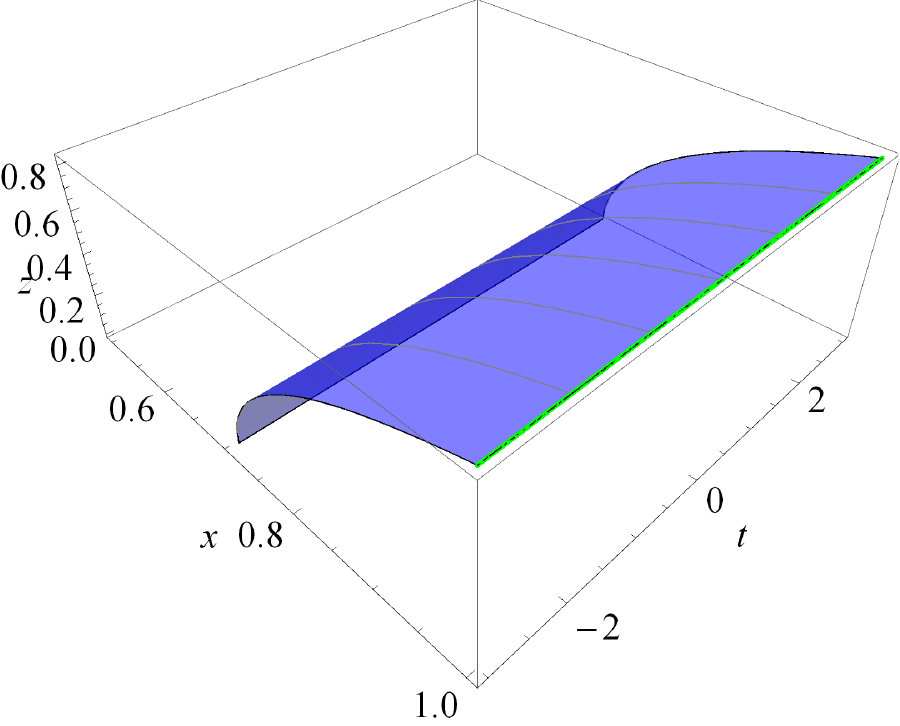}
\hskip 0.1in
\caption{The Lorentzian worldsheet as viewed by the EPR observer (left panel) and by the Hawking observer (right panel). In both plots we have set $\tilde{x}_0=0.7$. The worldsheet horizons are depicted in green, while constant time slices appear in gray. All quantities are measured in units of $H$ which we have set to unity. The quark trajectories can be obtained by truncating the embeddings at $z=z_m$. In both cases it is clear that we must impose $z_m<z^{ws}_{H}$ in order to have time-like trajectories. \label{staticobservers}}
\end{figure*}

The coordinate transformation to the static patch of the EPR observer is given by
\bea\label{statictoglobaltau}
& \displaystyle \tau  = \frac{1}{H}\mathrm{arcsinh}{(-\sqrt{1-H^{2}x^{2}}\sinh{(Ht)})}\,,\\
& \displaystyle \phi  = \mathrm{arctan}{\left(-\frac{Hx}{\sqrt{1-H^{2}x^{2}}}\mathrm{sech}{(Ht)}\right)}\,,\label{statictoglobalphi}
\eea
and the worldsheet translates to
\begin{equation}\label{eprws}
z =  \frac{2}{H}\sqrt{\frac{\sqrt{1-H^{2}x^{2}}\cosh{(Ht)}-\sqrt{1-H^{2}\tilde{x}_0^{2}}}{\sqrt{1-H^{2}x^{2}}\cosh{(Ht)}+\sqrt{1-H^{2}\tilde{x}_0^{2}}}}\,.
\end{equation}
The parameter $\tilde{x}_{0}$ is related to $\alpha$ through
\be
\tilde{x}_{0} = \frac{\sin{\alpha}}{H}\,,
\ee
and is related to the initial separation of the pair for the infinite massive case. For finite mass, the
the trajectory of the pair can be obtained by evaluating (\ref{eprws}) at $z=z_m$:
\be
x{(t)} = \pm \frac{\mathrm{sech}{(Ht)}}{H}\sqrt{H^{2}x_{0}^{2}+\sinh^{2}{Ht}}\,,
\ee
where we defined
\be\label{x0tildex0}
x_{0} = \frac{\sqrt{\tilde{x}_{0}^{2}(4+H^{2}z_{m}^{2})^{2}-16z_{m}^{2}}}{4-H^{2}z_{m}^{2}}\,.
\ee
The embedding (\ref{eprws}) is plotted in Figure \ref{staticobservers}. From the point of view of this observer, two particles come out of the horizon at $t\to-\infty$ reach a minimal value of $|x|$ at $t=0$ and finally fall back into the horizon at $t\to\infty$. This means that the EPR observer is in causal contact with the pair for all (static) time $t$.  Therefore, it inherits the causal structure of the super-observer (global coordinates) given that it has access to (part of) all four regions of the wormhole. The constant-$t$ profiles evolve from a semicircle (as in the flat space case) at $t=0$, and become $\sqcap$-shaped at $t\to\pm\infty$. In this limit, the worldsheet is delimited in the $x$ direction by the cosmological horizon of the EPR observer, located at $x=\pm1/H$, and in the $z$ direction by the bulk horizon, $z=2/H$. In terms of $\tilde{x}_0$ the worldsheet horizon (\ref{wshorizon}) is given by
\be
z^{ws}_{H} = \frac{2}{H^{2}\tilde{x}_{0}}\left(1-\sqrt{1-H^{2}\tilde{x}_{0}^{2}}\right)\,,
\ee
and is shown in green in Figure \ref{staticobservers}. Also, from this plot it becomes clear the point we made at the end of the previous section regarding the possible range of $z_m$: if we truncate the worldsheet at $z_m>z^{ws}_{H}$ the trajectory of the pair would be space-like and, therefore, would not correspond to a situation of physical relevance.

Similarly, we can transform the Lorentzian worldsheet into the static patch coordinates of the Hawking observer. This can be easily achieved if we keep in mind that $\phi$-coordinate of the Hawking observer differs from that of the EPR observer only by a shift of $\frac{\pi}{2}$. After the transformation, the worldsheet embedding becomes:
\begin{equation}\label{Hawkingobserver}
z= \frac{2}{H}\sqrt{\frac{x-\tilde{x}_{0}}{x+\tilde{x}_{0}}}\,,
\end{equation}
where, in this case
\be
\tilde{x}_{0} = \frac{\cos{\alpha}}{H}\,.
\ee
We have actually recovered the disconnected solution (\ref{Lorentziandisconnected}) for an isolated static quark! At first glance this may seem surprising, since the calculation in this section is set up for an accelerating pair of particles, rather than the ``static'' situation described in Section \ref{sec:potential}. However, it is important to bear in mind that, any particle who remains at constant coordinate $x$ is actually undergoing constant acceleration toward the geodesic observer at the center (see equation (\ref{acceqn})). Notice that, from the point of view of this observer, the electric field needed to sustain the quark's worldline actually compensates for the gravitational repulsion due to the de Sitter space. The position of the quark is obtained from the embedding (\ref{Hawkingobserver}):
\be\label{x0tilde}
x(t)=x_0=\tilde{x}_{0}\left(\frac{4+H^{2}z_m^{2}}{4-H^{2}z_m^{2}}\right)\,,
\ee
and the worldsheet horizon is now at
\begin{equation}
z^{ws}_{H} = \frac{2}{H}\sqrt{\frac{1-H\tilde{x}_0}{1+H\tilde{x}_0}}\,.
\end{equation}
It is clear that, although the Hawking observer does not have access to all four regions of the wormhole, the ``maximally extended'' version of the worlsheet (\ref{Hawkingobserver}) is the same as in the global case, since we are dealing with a simple coordinate transformation. This provides supporting evidence for the ER$=$EPR conjecture, now applied to Hawking pairs. The embedding (\ref{Hawkingobserver}) is plotted in Figure \ref{staticobservers}. As we can see, the worldsheet horizon is located exactly where the worldsheet intersects the cosmological horizon, $x=1/H$. This implies that $z_m$ is also constrained to be $z_m<z^{ws}_{H}$ since, from the point of view of this observer, the worldsheet actually terminates at $z=z^{ws}_{H}$.

\section{Summary and conclusions\label{sec:discussion}}

In this paper, we have analyzed the Schwinger mechanism in de Sitter space from the holographic viewpoint. Even though this effect has been previously studied in the literature, most of the existing results were derived using standard field theoretical methods which are valid for the weak coupling and weak electric field regimes. The general consensus suggests the existence of an upper critical value for the electric field for which the potential barrier for pair creation disappears, rendering the vacuum catastrophically unstable. However, such value is parametrically so large that it lies outside the regime of validity of the perturbative approximation. One of the main goals of this work was to investigate the existence of such a critical electric field from a non-perturbative point of view and, in order to achieve this, we have used various tools of the AdS/CFT correspondence. In the following, we will briefly summarize the most important lessons of our work.

In the first part of this paper, in Section \ref{sec:potential}, we performed a potential analysis of the Schwinger mechanism in de Sitter space. The computation was carried out in two steps. In Section \ref{subqqbar} we first considered a pair of infinitely massive particles in the absence of any electric field. In order to obtain the binding energy of the pair we computed the expectation value of certain Wilson loop
operator. We focussed on the static patch of de Sitter since, in this case, the energy of the pair is well defined. The final result is shown in Figure \ref{Potrectangular}. As expected, the short distance behavior of the potential (\ref{qqsmall}) reduces to the standard Coulombic profile. However, the full potential receives a series of corrections that become important at large distances. This should not be surprising since, in de Sitter space, we introduce a lengthscale $H^{-1}$ into the theory and, therefore, the potential is expected to contain additional terms that can be expressed in powers of the dimensionless quantity $H\ell$. In addition, we found that the system undergoes a first order phase transition at $H\ell_{\text{scr}}\approx0.75$. For larger distances the two particles are screened by the bath of Hawking quanta, in complete analogy to the phenomenon of Debye screening in classical electromagnetism. Notice that a similar effect has also been found in other cosmological setups \cite{Barosi:2008cs}.
In section \ref{subeffqqbar} we generalized the previous result in two ways: we included finite mass quarks and we turned on a background electric field. The effective potential in this case is plotted in Figure \ref{fig:Schwinger_barriers}, for various values of the electric field. From this computation, we were able to obtain the critical value of the electric field, by studying the short distance behavior of the potential (\ref{eq:veffEc}) and demanding that the tunneling barrier disappears at $E=E_c$. Our result shows that the value of the critical field is smaller than its flat space counterpart, implying that the vacuum is less stable in de Sitter space than in Minkowski space. This can be intuitively understood since the
expansion of the universe is itself a source of particle creation, thus, lowering the destabilizing threshold for the electric field.

In Section \ref{sec:nucleation}, the Schwinger effect was reanalyzed by explicitly constructing the Euclidean instanton for tunneling through the potential barrier of pair creation. Given the symmetries of the problem, the instanton is given by a circular loop that wraps $n$ times a sphere (the Euclidean continuation of de Sitter) at a constant polar angle $\theta_0$. The holographic dual of the instanton is found to be a spherical cap living in the Poincar\'e ball (the Euclidean continuation of the bulk geometry) subtended by a polar angle $\theta_0$ and, thus, intersecting the boundary at the same loop describing the instanton. The nucleation rate, $\Gamma\sim e^{-S_{E}}$, is computed from the area of the spherical cap and extremizing with respect of the angle $\theta_0$. The final result is a little cumbersome but in the regime $m\gg H\sqrt{\lambda}$ the expression reduces significantly (\ref{productionrate}). From this expression we can easily identify three contributions: the first term is the usual Boltzmann
factor with the de Sitter temperature, $T_{\text{dS}}=H/2\pi$, and therefore is the production rate due to the expansion of
space. The second term is linear in the electric field and agrees with the expected result from weakly coupled computations \cite{Garriga:1993fh}.
In addition, there are terms containing higher powers of the electric field. From (\ref{NGSE}), it is clear that those terms arise from the Wilson loop insertion and, therefore, can be interpreted as non-perturbative contributions to the nucleation rate. Our result in this regime also agrees with the semi-classical analysis presented in \cite{Frob:2014zka}. Finally, it is worth pointing out that the computation of the nucleation rate confirms the value for the critical electric field obtained in Section \ref{sec:potential}. It is easy to check that in the limit $E\to E_c$ the Euclidean action $S_E$ of the instanton vanishes. This implies that when this critical value is reached, the instanton sum is no longer exponentially suppressed, therefore, rendering the vacuum unstable.

In the last part of this work, in Section \ref{sec:EPR}, we analytically continued the Euclidean worldsheet of the tunneling instanton and studied its causal structure. We showed that the induced geometry on the worldsheet resembles the dynamical creation of a wormhole which, as explained, provides further evidence in support of the ER$=$EPR conjecture. We emphasize that this is a feature of the \emph{worldsheet} geometry (which represents the flux tube connecting the members of the EPR pair), rather than the bulk geometry itself, so the ER bridge should be thought of as a `gluonic' wormhole subtended by the pair, as opposed to a `spacetime' or `gravitational' wormhole. On the other hand, since no observer in de Sitter space has access to the entire manifold, we specialized to the case of different static observers. We focused on two special cases, an ``EPR observer'' and a ``Hawking observer'', both depicted in Figure \ref{staticobservers}. From the point of view of the EPR observer the two particles always lie inside its horizon. Therefore, the two particles in this case are interpreted as an usual EPR pair
nucleated from the de Sitter vacuum. The Hawking observer, on the other hand, only has causal contact with one of the particles, while the other always lies behind its horizon. For this observer, the particles are interpreted as a Hawking pair nucleated from the cosmological horizon.
We conclude that, regardless the observer point of view, the ER$=$EPR interpretation holds in a similar way since in all cases the causal structure of the worlsheet is inherited from that of the super-observer.

\section*{Acknowledgements}

The authors acknowledge useful conversations and correspondence with Alberto G\"uijosa, Takeshi Kobayashi and Diego Trancanelli. This material is based upon work supported by the National Science Foundation under Grant Number PHY-1316033 and by Texas Cosmology Center, which is supported by the College of Natural Sciences and the Department of Astronomy at the University of Texas at Austin and the McDonald Observatory. JFP also acknowledges support from Perimeter Institute for Theoretical Physics. Research at Perimeter Institute is funded by the Government of Canada through Industry Canada and by the Province of Ontario through the Ministry of Research and Innovation.

\appendix

\section{Maxwell's equations\label{AppEF}}

\begin{figure}
\centering
\includegraphics[width=4cm]{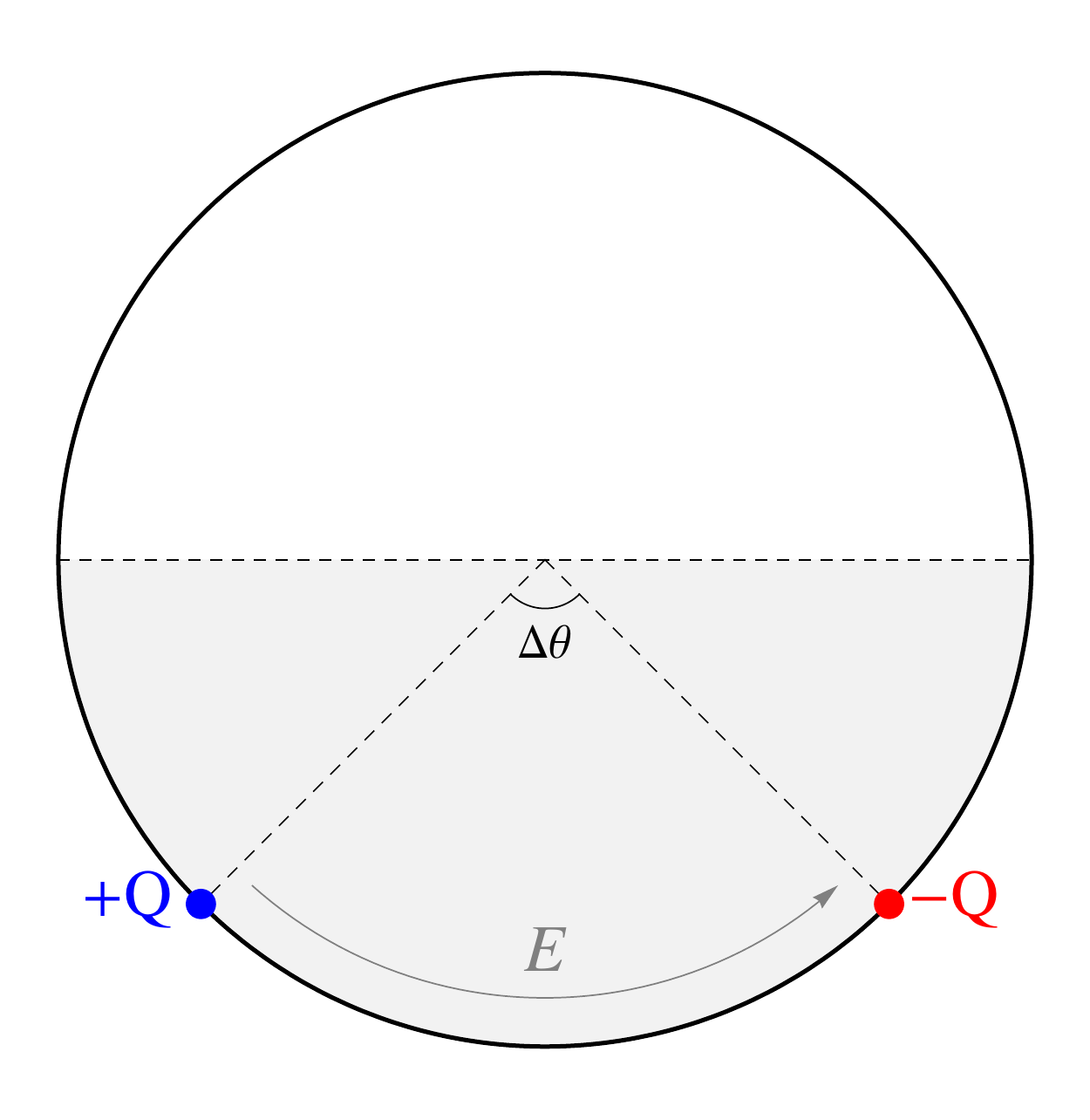}
%\hskip 1cm
\includegraphics[width=4cm]{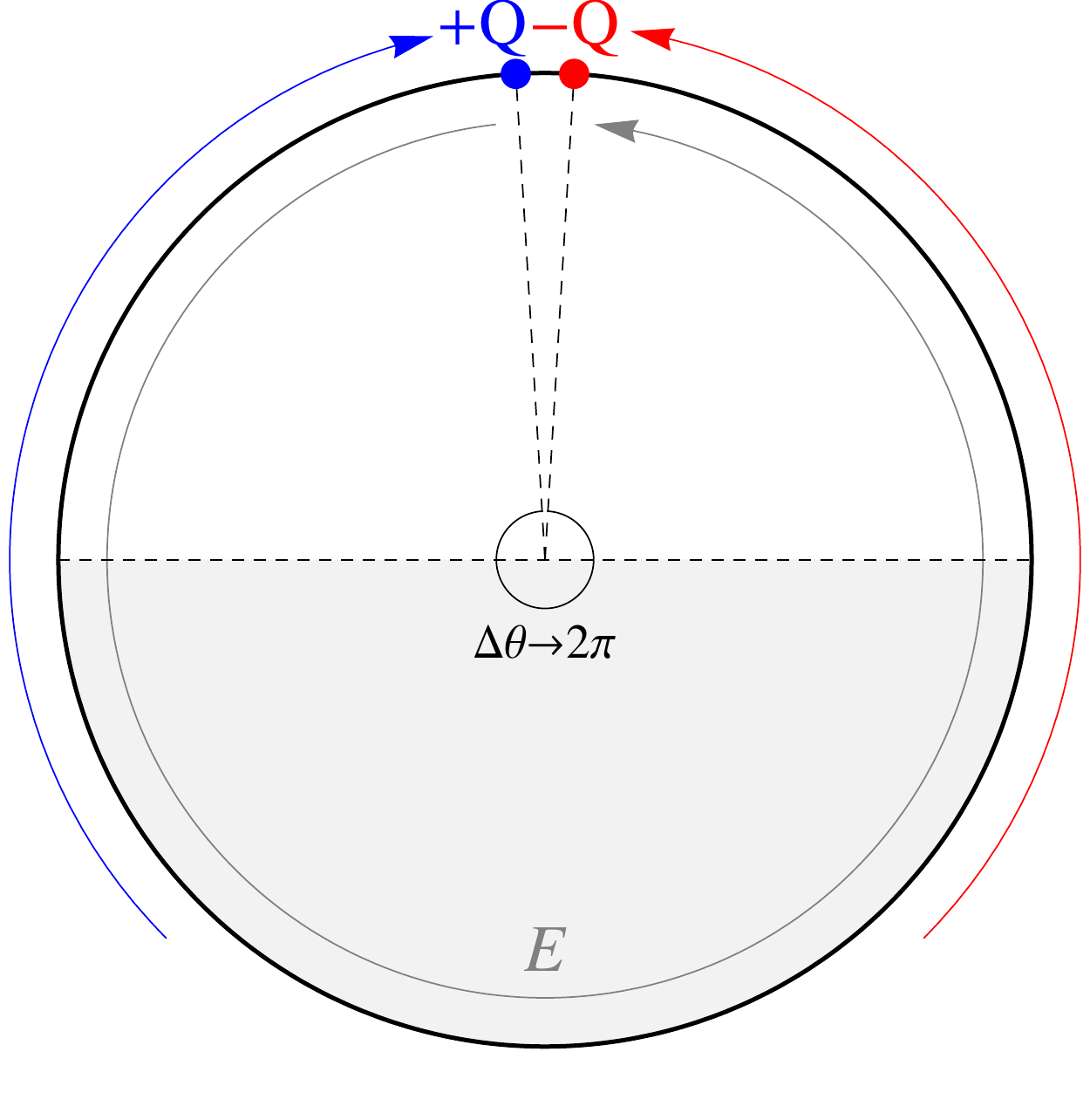}
\caption{Setup for a constant electric field in de Sitter space in $1+1$ dimensions. Two charges $+Q$ and $-Q$ separated by an angle $\Delta\theta$ are placed in the de Sitter manifold. If $\Delta\theta<\pi$ the two charges are visible to a static observer equidistant to the two particles. The portion of de Sitter accessible to such an observer consists of  the lower half of the circle and is delimited by a horizon located at $\theta=\{-\pi/2,\pi/2\}$. If $\Delta\theta>\pi$ the static observer will detect a constant electric field due to smeared charges at the horizon. Finally, if we let $\Delta\theta\to2\pi$ the two charges effectively cancel out, leaving us with a constant electric field in the entire manifold with no charges.\label{charges2D}}
\end{figure}

Let us take a closer look at Maxwell's equations in de Sitter space, focusing for now on the static patch of de Sitter (\ref{staticdS}). For a constant electric field in $1+1$ dimensions we have
\be
F_{\mu\nu}=\left(
             \begin{array}{cc}
               0 & E \\
               -E & 0 \\
             \end{array}
           \right)\,,
\ee
where $x^\mu=(t,x)$ are the coordinates of a static observer. There are many worldlines in the patch that have constant proper acceleration. For concreteness let us consider a particle sitting at a constant-$x$ orbit, with 4-velocity given by
\be
u^\mu=\left(\frac{1}{\sqrt{1-H^2x_0^2}},0\right)\,.
\ee
For a test particle with unit charge, the geodesic equation is satisfied as long as the Lorentz force cancels exactly with the repulsion due to de Sitter space, \emph{i.e.}
\be
a^\mu\equiv\frac{d u^\mu}{d\tau}+\Gamma^{\mu}_{\alpha\beta}u^\alpha u^\beta=F^\mu_\nu u^{\nu}\,.
\ee
In particular, this implies the following relation between the position $x_0$ and the electric field $E$:
\be\label{Efieldxo}
Hx_0=\frac{E}{\sqrt{E^2+H^2}}\,.
\ee
The magnitude of the acceleration is constant, as it should be, and it is determined by the particle's position $x_0$ through:
\be\label{a2x0Erel}
a^2\equiv a_\mu a^\mu=\frac{H^4x_0^2}{1-H^2x_0^2}=E^2\,.
\ee
A charged particle located at $x>x_0$ will accelerate towards the horizon. On the other hand, if we place it at $x<x_0$ the particle will accelerate away from the horizon. It is easy to check that Maxwell's equations are trivially satisfied without sources,
\be
\nabla_\mu F^{\mu\nu}=0\,.
\ee
To understand the origin of this electric field we can define a global coordinate $\theta$ through $Hx=\sin \theta$. The spatial part of the metric becomes that of a circle of radius $H^{-1}$. However, it is clear that the static patch only covers half of it, from $\theta=-\pi/2$ to $\theta=\pi/2$. To achieve a constant electric field in the static patch we can place a ``capacitor'' consisting of two charges $+Q$ and $-Q$, located at $\theta=-\theta_0$ and $\theta_0$, respectively. See figure \ref{charges2D} for a schematic representation. If the separation is such that $\Delta\theta\equiv2\theta_0>\pi$, the static observer will see the charges ``smeared'' at the horizon and still detect the presence of the electric field. Finally, if we let $\Delta\theta\to2\pi$, the two charges overlap and effectively cancel out. This implies that a constant electric field can be achieved in global de Sitter without the addition of any source.

\begin{figure}
\centering
\hskip -1cm
\includegraphics[width=8cm]{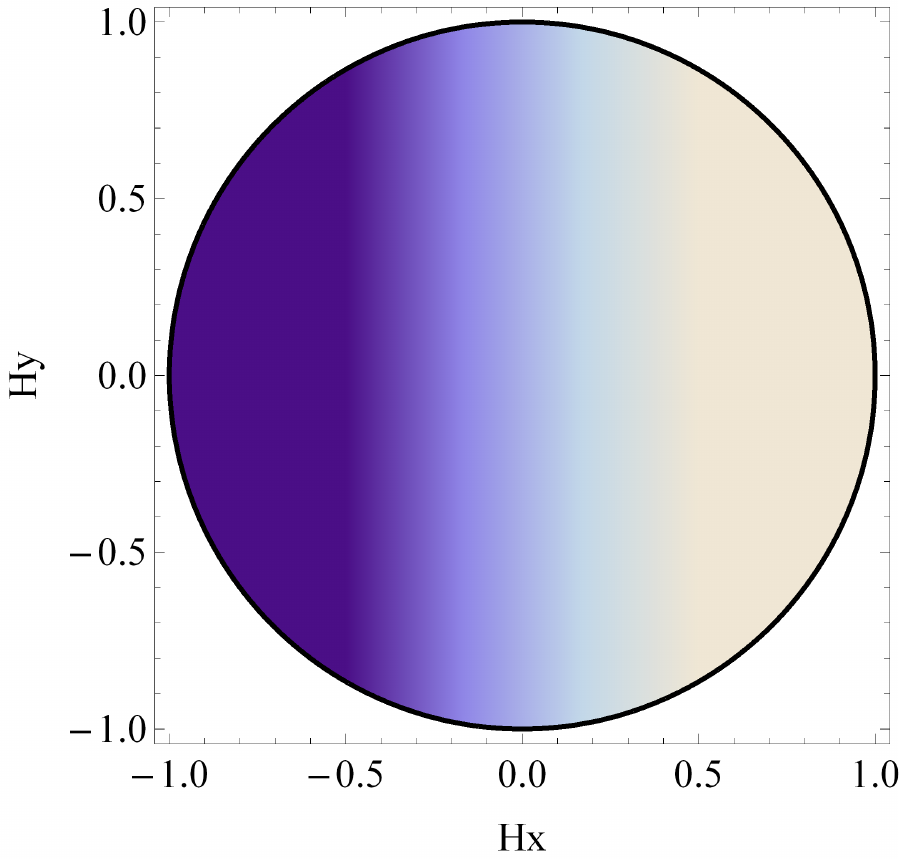}
\caption{Density plot for $J_t/E H^2$ in the equatorial plane $\theta=\pi/2$, assuming a constant electric field along the $x$-direction, $F_{tx}=E$. The total charge in the static patch evaluates to zero, but the distribution breaks the $SO(3)$ symmetry.\label{charge4D}}
\end{figure}

Unfortunately, the above analysis does not hold in higher dimensional de Sitter space. In the following we will illustrate the similarities and differences for the specific case of $3+1$ dimensions. Let us start by considering a constant electric field along one of the space directions, say $F_{tx}=-F_{xt}=E$, and all other components turned off. In spherical coordinates $(r,\theta,\phi)$ the non-zero components of the field strength are given by:
\bea
&&F_{tr}=F_{rt}=E \sin\theta \cos\phi\,,\nonumber\\
&&F_{t\theta}=F_{\theta t}= E\, r \cos\theta \cos\phi \,,\\
&&F_{t\phi}=F_{\phi t}=- E\, r \sin\theta \sin\phi\,.\nonumber
\eea
There are many trajectories in the static patch that have constant proper acceleration. For concreteness, however, we will focus on 
worldlines that are simple generalizations of the $1+1$ dimensional case. We will assume that the particle lies on the $y=z=0$ plane and follows a constant-$x$ orbit. Therefore, its 4-velocity is given by
\be
u^\mu=\left(\frac{1}{\sqrt{1-H^2x_0^2}},0,0,0\right)\,.
\ee
It is straightforward to show that both (\ref{Efieldxo}) and (\ref{a2x0Erel}) hold in this case, so by tuning the value of $E$ we can achieve the desired trajectory with constant acceleration. However, Maxwell's equations implies now that
\be
\nabla_\mu F^{\mu\nu}=J^\nu\,,
\ee
with
\be
J^{\mu}=\left(-\frac{2 E H^2 r \sin\theta \cos\phi}{1-H^2 r^2},0,0,0\right)\,.
\ee
This implies that in order to source the desired electric field, we must have a charge density:
\be
J_t\equiv\rho(r,\theta,\phi)=2 E H^2 r \sin\theta\cos\phi=2 E H^2 x\,.
\ee
The total charge in the static patch evaluates to zero, but the distribution explicitly breaks the $SO(3)$ symmetry (see figure \ref{charge4D}). Moreover, it is clear that although the electric and gravitational interactions are in balance, the charge configuration is classically unstable.

\section{The energy of an isolated quark\label{AppSingleQ}}

In this appendix we obtain the string embedding corresponding to an isolated static quark in de Sitter space. Then, we employ the solution in order to compute its energy. The final result can be expressed as the sum of two contributions, the rest mass of the quark plus the effective gravitational potential energy which is a function of the position.

For a single quark located at $x=0$ the solution for the embedding corresponds to a vertical string that stretches between the boundary and the bulk horizon \cite{Fischler:2014tka}. However, if we place the quark at some finite distance from the center $x=x_0$ the string is no longer vertical but bends towards the de Sitter horizon (located at $x=1/H$). This is just a reflection of the fact that a purely vertical embedding would not have minimal area: just like in the boundary theory trajectories at fixed $x_0>0$ are non-geodesic, objects in the bulk experience a gravitational force that pushes away from the origin.

The situation is very similar to the case in Rindler space, in which the solution for the string embedding is bent and follows a path of minimal proper length towards the acceleration horizon \cite{Caceres:2010rm}. With this intuition in mind, let us first consider space-like geodesics in the bulk geometry. If we choose the affine parameter to be $\lambda=x$, the proper length of a curve $z(x)$ is given by
\be\label{geoaction}
\mathcal{S}=\int ds=L\int\frac{dx}{z}\sqrt{\frac{f(z)^2}{h(x)}+z'^2}\,,
\ee
where $f(z)$ and $h(x)$ are the functions defined in (\ref{fz}) and (\ref{hx}), respectively. From (\ref{geoaction}) we can derive the equation of motion for the geodesics:
\bea\label{eomgeo}
&&\!\!\!\!\!z''+\frac{h'(x)}{2h(x)}z'+\frac{ f(z)-2 z f_z(z)}{z f(z)}z'^2\nonumber\\
&&\qquad\qquad\qquad\quad\;\;+\frac{f(z) (f(z)-z f_z(z))}{z h(x)}=0\,,
\eea
with $f_z(z)\equiv\partial_zf(z)$. Albeit non-linear, equation (\ref{eomgeo}) can be solved analytically. The general solution can be written in terms of two integration constants $A$ and $B$:
\be\label{geosol}
z(x)=\frac{2}{H} \sqrt{\frac{A \sin \left(B+\arccos(H x)\right)-1}{A \sin \left(B+\arccos(H x)\right)+1}}\,.
\ee
Imposing that $z(x_0)=0$ fixes one of the constants in terms of the other, \emph{e.g.},
\be\label{consgeo}
A=\csc \left(B+\arccos(H x_0)\right)\,.
\ee
Thus, at this point we have a solution depending on a single parameter $B$. Depending on its value, the geodesic might go back to the boundary at some other $x=x_f$ or end up hitting the de Sitter horizon at $x=1/H$ and some $z=z_f$.

We claim that the solution for the string embedding representing an isolated quark at $x=x_0>0$ follows a path of minimal length towards the de Sitter horizon. Indeed, we can explicitly check this by plugging (\ref{geosol}) into the equation of motion of the string (\ref{eom}). This picks up a unique physical solution with
\be
B=\frac{\pi}{2}\quad (\text{mod}\; \pi)\,,
\ee
and profile
\be
z(x)=\frac{2}{H}\sqrt{\frac{x-x_0}{x+x_0}}\,.\label{zofxdisconnected}
\ee
This result should not be surprising. The role of spacetime geodesics for reconstructing string and brane embeddings was recently pointed out in \cite{Fiol:2014vqa}. We expect a similar construction to be possible even for dynamical configurations.

With this solution at hand, we can now compute the energy of the quark as a function of $x_0$. To do this, we need to compute the conjugate momentum of the Nambu-Goto action.\footnote{For a review on quark dynamics in AdS/CFT see \cite{Chernicoff:2011xv}.} Before gauge-fixing, the worldsheet is parametrized by functions of $\tau$ and $\sigma$, and the momentum densities are given by
\begin{equation}\label{energydens}
\Pi_{\mu} = \frac{\partial \mathcal{L}_{\text{NG}}}{\partial \dot{X}^{\mu}} = \frac{1}{2\pi\alpha'}\frac{\dot{X}_{\mu}X'^{2}-X_{\mu}'(\dot{X}\!\cdot\! X')}{\sqrt{\dot{X}^{2}X'^{2}-(\dot{X}\!\cdot\! X')^{2}}}\,,
\end{equation}
where $\dot{\,}\equiv\partial_\tau$, $'\equiv\partial_\sigma$ and $\dot{X}\!\cdot\!X'=G_{\mu\nu}\dot{X}^{\mu}X'^{\nu}$. In particular, the energy density is the time component of (\ref{energydens}). To compute the energy we choose to work in the gauge $(\tau,\sigma)=(t,z)$, so that $X^{\mu} = (t,x(z),z)$.
In this gauge the energy density takes the form
\begin{equation}\label{energydensity}
\mathcal{E} = \frac{\sqrt{\lambda}}{2\pi}\frac{f(z)}{z^2}\sqrt{h(x)+f(z)^2x'^{2}}\,,
\end{equation}
which is, indeed, equivalent to the Nambu-Goto Lagrangian $\mathcal{L}_{\text{NG}}$.
The function $x(z)$ can be obtained by inverting equation (\ref{zofxdisconnected}), and takes the following form:
\begin{equation}\label{solxinv}
x{(z)} = \tilde{x}_{0}\left(\frac{4+H^{2}z^{2}}{4-H^{2}z^{2}}\right)\,.
\end{equation}
Notice that we have relabeled the parameter $x_0\to \tilde{x}_0$. The reason for this is that,
for the case of finite mass, the string actually ends at a fixed bulk depth $z_m$ (where the flavor branes are located) and the embedding of interest is just the $z \geq z_m$ portion of the solution (\ref{zofxdisconnected}). The parameter
$\tilde{x}_0$ (now an auxiliary variable) is related to the physical $x_0$ through
\be\label{x0tilde}
x_0=\tilde{x}_{0}\left(\frac{4+H^{2}z_m^{2}}{4-H^{2}z_m^{2}}\right)\,,
\ee
and substituting in (\ref{energydensity}) the latter becomes
\be\label{physicalx}
x{(z)} =x_0 \left(1+\frac{8 H^2(z^2-z_m^2)}{(4-H^2 z^2)(4+H^2 z_m^2)}\right)\,.
\ee
Notice that (\ref{physicalx}), in fact, satisfies the expected relation $x(z_m)=x_0$.
For this solution, the energy density (\ref{energydensity}) evaluates to
\begin{equation}
\mathcal{E} = \frac{\sqrt{\lambda}}{2\pi}\frac{f(z)}{z^2}\sqrt{1-H^{2}\tilde{x}_0^2}\,.
\end{equation}

Integrating from $z_{m}$ up to the maximum value of $z$,
\begin{equation}
z^{ws}_{H}\equiv z(1/H)= \frac{2}{H}\sqrt{\frac{1-H\tilde{x}_{0}}{1+H\tilde{x}_{0}}}\,,
\end{equation}
we find the total energy to be
\begin{equation}\label{totenergyq}
E=\frac{\sqrt{\lambda}}{2\pi z_m}\left[\left(1+\frac{H^2z_m^2}{4}\right)\sqrt{1-H^{2}\tilde{x}_0^2} - H z_m\right].
\end{equation}
This can be expressed as a sum of two contributions,
\be
E\equiv m+V_{\text{grav}}(x_0)\,,
\ee
where $m$ is the mass of the quark and $V_{\text{grav}}(x_0)$ is the effective gravitational potential energy. In particular, for $x_0=0$, we obtain
\be\label{massquark}
E=m =\frac{\sqrt{\lambda}}{2\pi z_m}\left(1-\frac{Hz_m}{2}\right)^2\,.
\ee
Notice that (\ref{massquark}) blows up as $z_m\to0$, as anticipated, and can be inverted to obtain
\be
\begin{split}\label{zminverted}
z_m=&\;\frac{\sqrt{\lambda }}{\pi  \left(m+m \sqrt{1+\frac{\sqrt{\lambda }H}{\pi  m}}+\frac{\sqrt{\lambda }H}{2 \pi }\right)}\,,\\
=&\;\frac{\sqrt{\lambda }}{2 \pi  m}\left[1-\frac{\sqrt{\lambda }H}{2 \pi  m}+\mathcal{O}\left(\frac{\lambda H^2}{m^2}\right)\right]\,.
\end{split}
\ee
We stress, however, that we are only allowed to treat the string semiclassically as long as it is sufficiently heavy. This means that we are restricted to work in the regime $z_m\ll z_H$ or, equivalently, $m\gg \sqrt{\lambda}H$. For a neater interpretation, we could alternatively split (\ref{massquark}) in two pieces, by noticing that the $H$-dependent terms should be regarded as the thermal correction to the mass \cite{Fischler:2014tka}
\be
m=m_0-\delta m_H\,,
\ee
where
\be
m_0=\frac{\sqrt{\lambda}}{2\pi z_m}\,,\qquad\delta m_H=\frac{\sqrt{\lambda }H}{2 \pi } \left(1-\frac{H z_m}{4}\right)\,,\nonumber
\ee
and
\be
\frac{\delta m_H}{m_0}=\frac{\sqrt{\lambda }H}{2 \pi  m}+\mathcal{O}\left(\frac{\lambda H^2}{m^2}\right)\ll1\,.
\ee

The effective gravitational potential can be obtained by subtracting (\ref{massquark}) from (\ref{totenergyq}) and is found to be
\be\label{effgravV}
V_{\text{grav}}(x_0)=\frac{\sqrt{\lambda }}{2 \pi  z_m}\left(1+\frac{H^2 z_m^2}{4}\right)\left(\sqrt{1-H^2 \tilde{x}_0^2}-1\right) \,.
\ee
We emphasize the word ``effective'' in the above definition. Physically, (\ref{effgravV}) includes both, the ``bare'' gravitational energy $V_{\text{grav}}^{\text{bare}}(x_0)$ plus the self-energy $\Sigma(x_0)$ which also depends on the position and is due to due to interactions between the quark and the quantum fields in de Sitter.\footnote{The term $\delta m_H$ is also part of $\Sigma(x_0)$, but we choose to write it as a separate contribution so that $V_{\text{grav}}(0)=0$.}$^,$\footnote{The self-energy contribution is responsible for the peculiar behavior of the rms displacement of a quark which, at late times, is found saturate to a finite distance inside the horizon \cite{Fischler:2014tka}.} Using (\ref{x0tilde}) and (\ref{zminverted}) we could also write (\ref{effgravV}) in terms of the physical data, $m$ and $x_0$, but the result is not particularly illuminating. For small distances $H x_0\ll1$, however, the result takes the following form:
\be
V_{\text{grav}}(x_0)\approx-\frac{1}{2} m_0 H^2 x_0^2 \left[1+\mathcal{O}\left(\frac{\delta m_H}{m_0}\right)\right]\,.
\ee
The inverted harmonic potential is expected for a test particle in de Sitter space \cite{Fischler:2014tka}. The $\mathcal{O}(\delta m_H/m_0)$ correction, on the other hand, appears from the self-energy itself and should be thought of as a quantum effect.

\section{Analytic continuation}\label{App2}

\subsection{Euclidean de Sitter and Anti de Sitter\label{subsec51}}

Consider the AdS metric foliated with dS slices (\ref{eq:metric1}), and let us write the dS part in global coordinates,
\be
ds_{\text{dS}}^2=-d\tau^2+\frac{1}{H^2}\cosh^2(H\tau) d\phi^2\,.
\ee
We will also redefine the bulk coordinate according to
\begin{equation}\label{eqnB2}
z = \frac{2}{H}e^{-2\,\text{arctanh}\,r}\,.
\end{equation}
so that $r$ ranges from $0$ (for $z=2/H$) to $1$ (for $z=0$). To obtain the Euclidean counterpart of such geometry, we analytically continue the time coordinate  $\tau = i\frac{\tau_{E}}{H}$ both in the boundary and in the bulk. The resulting geometry is that of the Poincar\'e ball,
\begin{equation}\label{Poincareball}
ds^{2} = \frac{4L^{2}}{(1-r^{2})^{2}}\left[dr^{2} + r^{2}\left(d\tau_{E}^{2} + \cos^{2}{\tau_{E}}d\phi^{2}\right)\right]\,,
\end{equation}
where the Euclidean time is now identified as an angle.
This form of Euclidean AdS can be understood as follows: one starts with a 3-dimensional Euclidean space $\mathbf{R}^{3}$ with spherical coordinates $(r,\tau_{E},\phi)$, and puts the above metric on the ball $r<1$. The Poincar\'{e} ball is the unique Euclidean metric whose boundary is $\mathbf{S}^{2}$. This shows that the requirement of de Sitter invariance uniquely determines the bulk geometry.

Alternatively, we could have started with the static patch of dS instead of global dS. In this case, the de Sitter metric is given by
\begin{equation}\label{staticpatchdS}
ds_{\text{dS}}^{2} = -(1-H^{2}x^{2})dt^{2} + \frac{dx^{2}}{1-H^{2}x^{2}}\,.
\end{equation}
Redefining the $x$ coordinate according to
\begin{equation}\label{thetatostaticpatch}
Hx = \sin{\theta}\,,
\end{equation}
and analytically continuing the bulk metric, we arrive to
\begin{equation}\label{PoincareballSTATIC}
ds^{2} = \frac{4L^{2}}{(1-r^{2})^{2}}\left[dr^{2} + r^{2}\left(\cos^2\theta dt_{E}^{2} + d\theta^{2}\right)\right]\,,
\end{equation}
\emph{i.e.}, the boundary metric is also that of a sphere. We have to be careful about the global topology of the Euclidean space, however. Since the spacetime accessible to any single geodesic observer is only a fraction of de Sitter space this should be reflected in the analytic continuation. In fact, the correct topology of the Euclidean static patch is a hemisphere rather than the whole sphere. This means that for each static observer there is a different analytic continuation (in Euclidean signature, this corresponds to rotating the hemisphere around the sphere). Finally, we point out the fact that, in the line element (\ref{PoincareballSTATIC}), the Euclidean time $t_E$ plays the role of an azimuthal angle rather than a polar angle (as in the global de Sitter case).

\begin{figure}[t]
\includegraphics[width=7cm]{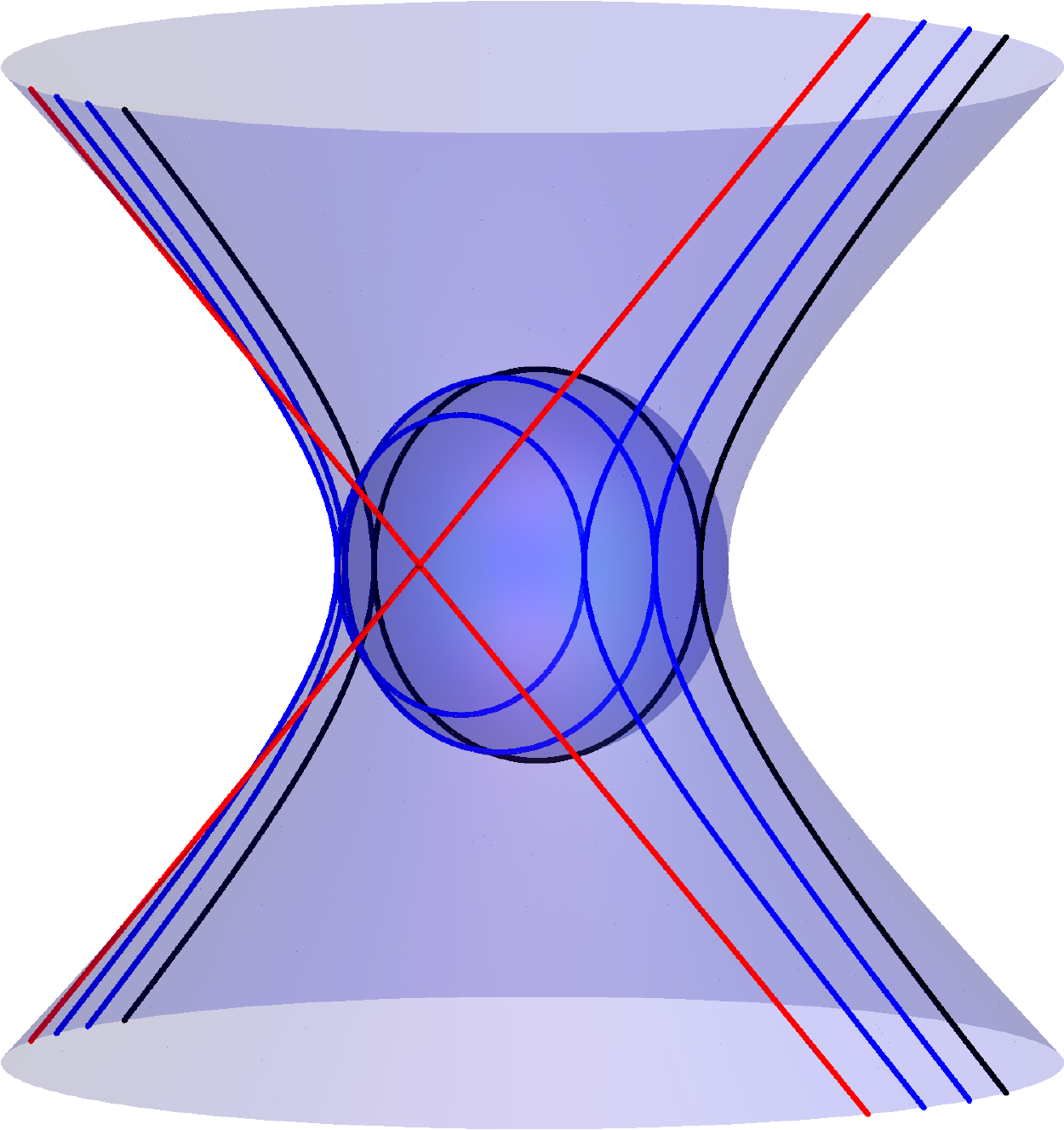}
\caption{The de Sitter manifold (as an hyperboloid embedded in Minkowski space) and its Euclidean counterpart (the sphere). Trajectories of constant proper acceleration (shown in blue) are obtained by intersecting the hyperboloid with vertical planes, and they become circular loops on the sphere. If the vertical plane also contains the origin of the ambient Minkowski space, then the trajectory is a geodesic (shown in black), and its Euclidean counterpart is a great circle.\label{EuclideandS}}
\end{figure}

Let us focus now on the Wilson loops we want to consider. It is easy to see that the worldlines needed for the computation of the quark-antiquark potential consist of two parallel circles with constant polar angles $ \pm \theta_0$. In Lorentzian signature, these trajectories can be obtained by intersecting an hyperboloid embedded in Minkowski space (this is de Sitter space by definition) with planes parallel to the time axis (see Figure \ref{EuclideandS}). A short calculation shows that, for these worldlines, the magnitude squared of the acceleration is given by
\begin{equation}\label{acceqn}
A^{2} = \frac{H^{4}x^{2}}{1-H^{2}x^{2}}\,.
\end{equation}
Thus, constant-$x$ trajectories are also trajectories with constant proper acceleration, with a magnitude that can vary from 0 for $x=0$ (that is, for a geodesic observer) to $\infty$ for $Hx \rightarrow 1$. Finally, notice that the Wilson loop needed for the nucleation rate corresponds also to one of such circles in Euclidean signature (given that these are precisely the worldlines with constant acceleration). To achieve this, we have to rotate the circle around the sphere, so that the static observer in consideration covers the portion of de Sitter that has access to the whole loop. After this transformation the circle is found to lie at a constant-$t_E$ surface. In Lorentzian signature, this means that the worldline will no longer be at constant-$x$ but, instead, will correspond to that of a pair of particles undergoing back-to-back uniform acceleration, as desired.

\subsection{Minimal surfaces}\label{Sectionminsurf}
In this section we describe the minimal surfaces necessary for the computation of the quark-antiquark potential and the nucleation rate. The results presented here are adapted from \cite{Krtous:2014pva}. The embedding functions are fully analytical, but are given in terms of a particular set of coordinates of Euclidean AdS.

\begin{figure*}[ht]
\centering
\includegraphics[width=5.8cm]{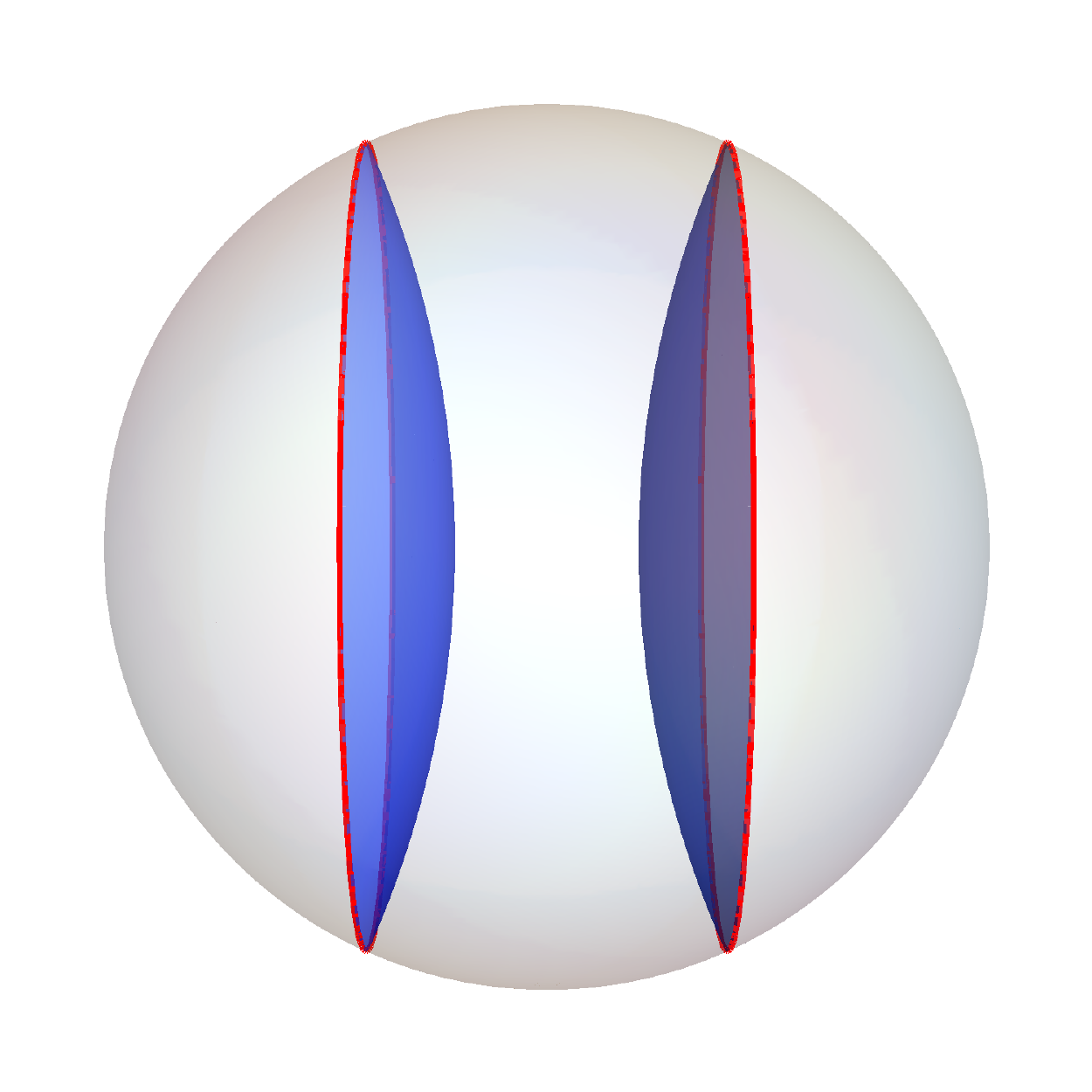}
\hskip 0.05in
\includegraphics[width=5.8cm]{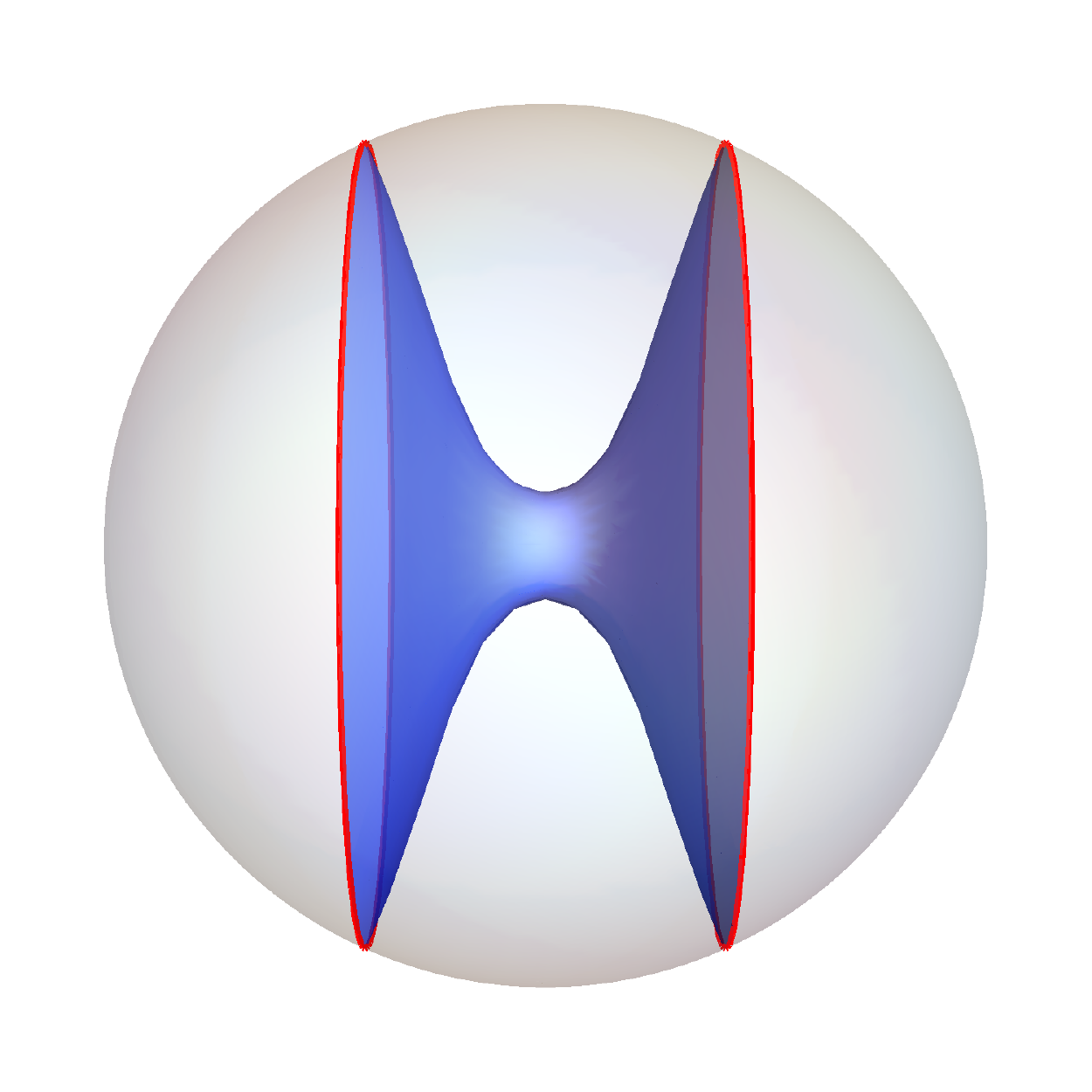}
\hskip 0.05in
\includegraphics[width=5.8cm]{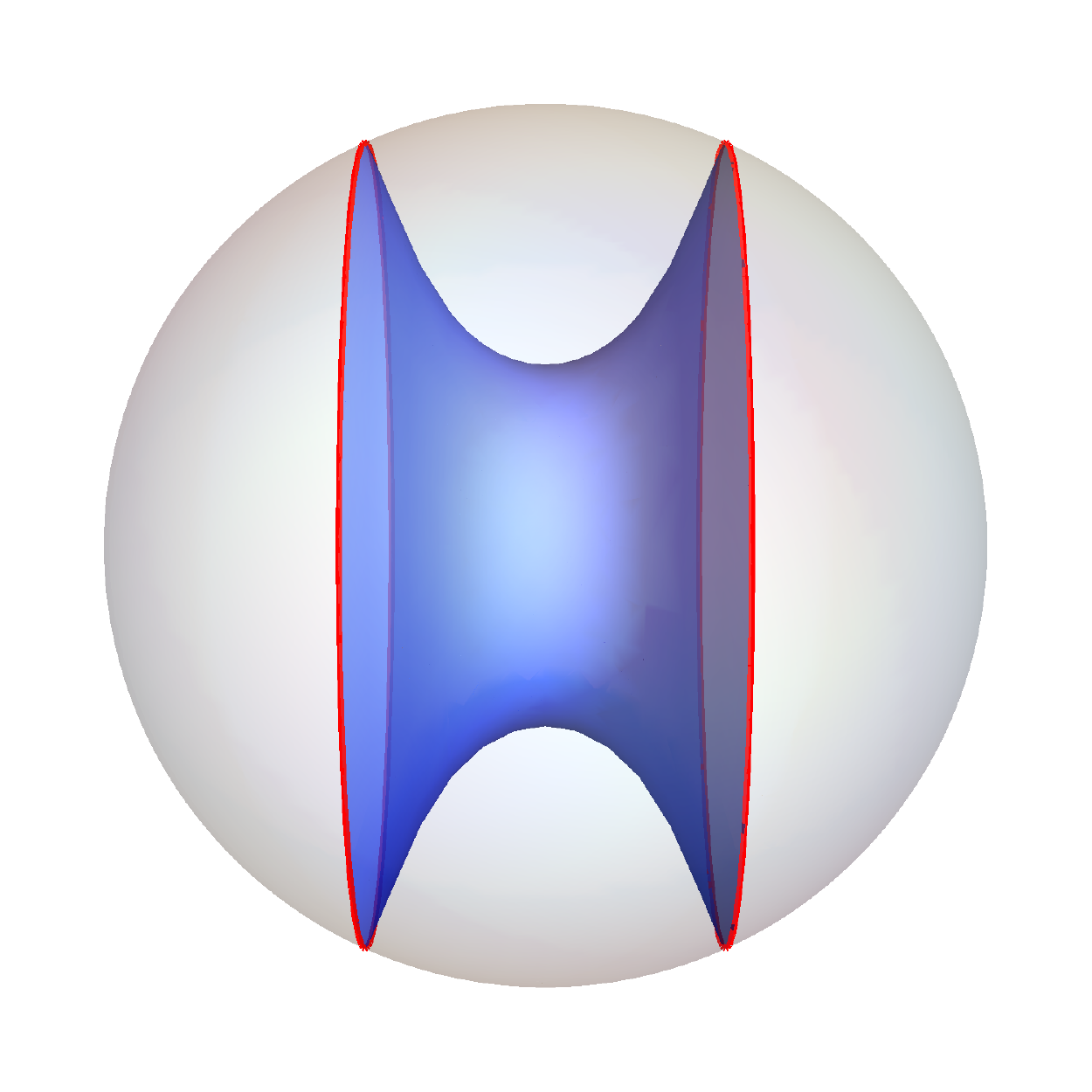}
\caption{The disconnected and the two connected extremal surfaces for a given separation, from the point of view of the Poincar\'e ball. \label{figconnected}}
 \end{figure*}

Euclidean AdS can be understood as a metric on a solid ball (the Poincar\'{e} ball), on a solid cylinder, or on a semi-infinite space (the Poincar\'{e} half-space), depending on the particular symmetry of the space that is emphasized. Of the three realizations, the Poincar\'{e} ball is the natural choice for our problem, because the Euclidean continuation of de Sitter space is a sphere. Now, as discussed in the previous section, we are interested in minimal surfaces that intersect the boundary at two parallel circles of the same size. There are in general three minimal surfaces satisfying the desired boundary condition: a disconnected one, and two connected ones. In Figure \ref{figconnected} we plot one example of each for a given separation.

Notice that the problem also has axial symmetry. As a matter of fact, it is simpler work in terms of the cylindrical parametrization first, and then transform the solutions back to the Poincar\'e ball \cite{Krtous:2014pva}. Euclidean AdS on the cylinder has a metric of the form
\begin{equation}\label{cylmetric}
ds^{2} =L^{2}\left[ \frac{dP^{2}}{1+P^{2}}+(1+P^{2})dZ^{2}+P^{2}d\varphi^{2}\right]\,.
\end{equation}
In this case, the boundary is defined as the cylinder at $P\to\infty$ and the two loops are just circles of constant $Z=\pm Z_\infty$. The coordinate transformation that relates (\ref{cylmetric}) to the Poincar\'{e} ball (\ref{Poincareball}) in the coordinate system $(r,\tau_{E},\phi)$ is given by
\bea\label{balltocylinderP}
&&P = \frac{2r}{1-r^{2}}\sqrt{1-\cos^{2}{\tau_{E}}\cos^{2}{\phi}}\,,\nonumber\\
\label{balltocylinderZ}
&&Z =\mathrm{arccosh}{\left(\frac{1+r^{2}}{\sqrt{(1+r^{2})^{2}-4r^{2}\cos^{2}{\tau_{E}}\cos^{2}{\phi}}}\right)}\,,\nonumber\\
&&\varphi = \arccos{\left(\frac{\cos{\tau_{E}}\sin{\phi}}{\sqrt{1-\cos^{2}{\tau_{E}}\cos^{2}{\phi}}}\right)}\nonumber\,.
\eea

Let us now discuss the solutions. The disconnected surface simply consists of two planes at constant $Z$ containing the loops:
\be
Z=\pm Z_\infty\,.
\ee
In the Poincar\'e ball, this solution can be written as:
\begin{equation}\label{discsolPB}
r=\frac{\cos{\phi}\cos{\tau_{E}}-\sqrt{\cos^{2}{\phi}\cos^{2}{\tau_{E}}-\cos^{2}{\alpha}}}{\cos{\alpha}}\,,
\end{equation}
with $\alpha=\frac{\pi}{2}\pm\theta_{0}$, and can be described as two spherical caps intersecting the boundary of the ball orthogonally. It is easy to see that (\ref{discsolPB}) indeed leads to the solution reported in (\ref{Lorentziandisconnected}) for the Lorentzian signature (see Appendix \ref{AppSingleQ} for an alternative derivation). The area of this surface (including both spherical caps) up to some cutoff value $P_m$ is given by:
\begin{equation}\label{areadisconnected}
A_{d} = 4\pi L^{2}(\sqrt{1+P_m^{2}}-1)\,.
\end{equation}

The connected surfaces are a little more complicated. In this case the solution
is given by
\be\label{connected}
Z(P) = \pm \gamma P_{0}\left[F{\left(\beta,\gamma\right)}
-\delta P_{0}^{2}\,\Pi\left(\beta,\delta,\gamma\right) \right]\,,
\ee
where
\bea\nonumber
\beta=\mathrm{arccos}\left(\frac{P_{0}}{P}\right)\,,\quad\gamma=\sqrt{\frac{1+P_{0}^{2}}{1+2P_{0}^{2}}}\,,\quad\delta=\frac{1}{1+P_{0}^{2}}\,.
\eea
 In the above, $F$ denotes the elliptic integral of the first kind and $\Pi$ denotes the complete elliptic integral of the third kind \cite{EllipticBook}. The parameter $P_{0}$ is an integration constant and corresponds to the smallest value of $P$ on the surface. The area of this surface up to some cutoff value $P_m$ is given by:
\begin{equation}\label{areaconnected}
A_{c}= \frac{4\pi L^{2}P_{0}^{2}}{\sqrt{1+2P_{0}^{2}}}\Pi\left(\beta_{m},1,\gamma\right)\,,
\end{equation}
where $\beta_m=\beta(P_m)$. The constant $P_m$ determines the value of $Z_m= Z(P_m)$ which, in the limit $P_m\to\infty$, reduces to $Z_\infty$. A brief calculation leads to
\be\label{Zinfinity}
Z_{\infty} = \gamma P_{0}\left[K{\left(\gamma\right)}-\delta P_{0}^{2}\,\Pi\left(\delta,\gamma\right) \right]\,,
\ee
where $K$ is the complete elliptic integral of the first kind and $\Pi$ is the incomplete elliptic integral of the third kind. Equation (\ref{Zinfinity}) is non-monotonic with respect to $P_0$. This implies that for a given boundary separation there are two minimal surfaces with different $P_0$, one reaching deeper into the bulk than the other (see Figure \ref{figconnected} for an example). Among these two, the solution with larger $P_0$ is the one with minimal area.

Finally, we will also need the renormalized area of the connected surface, $A_{r}=A_{c}-A_{d}$, which can be easily obtained from (\ref{areadisconnected}) and (\ref{areaconnected}). In the limit $P_m\to\infty$, in particular, this quantity simplifies to:
\be\label{renormalizedarea}
A_{r}^{\infty} = 4\pi L^{2}\left[1+\frac{P_{0}^{2}K{\left(\gamma\right)}}{\sqrt{1+2P_{0}^{2}}}
- \sqrt{1+2P_{0}^{2}}E{\left(\gamma\right)} \right],
\ee
where $E$ is the complete elliptic integral of the second kind.

\section{The nucleation rate in de Sitter space at weak coupling\label{App}}
In this appendix, we revisit the instanton computation of the QED production rate in de Sitter using the instanton method, that is, by solving the Euclidean equation of motion \cite{Garriga:1993fh}. %The results presented here will be compared against our result in the following subsection.
Let us start with the action for a charged particle in the presence of an electromagnetic field,
\begin{equation}
S = -m \int_{\gamma} \sqrt{-g_{\mu\nu}\dot{z}^{\mu}\dot{z}^{\nu}}d\tau + q\int_{\gamma} A\,,
\end{equation}
where $m$ and $q$ are the mass and charge of the particle, respectively, and $z^\mu=z^\mu(\tau)$. In Euclidean signature, the action becomes:
\begin{equation}\label{eucACT}
S_{E} = m \int_{\gamma} \sqrt{g_{\mu\nu}\dot{z}^{\mu}\dot{z}^{\nu}} d\tau + q\int_{\gamma} A\,.
\end{equation}
The contribution of an instanton to the path integral has the form $A e^{-S_{E}}$, where $A$ is a prefactor, and $S_{E}$ is the \emph{on-shell} Euclidean action. Thus, by computing $S_{E}$ we will recover the exponent in the Schwinger formula (\ref{Schwingerformula}).

Now, consider the case of a constant electric field where $F_{\mu\nu} = -E\sqrt{g}\epsilon_{\mu\nu}$. A brief computation shows that the trajectory that extremizes the action is a circle (see for example Figure \ref{EuclideandS}). Furthermore, since the worldline is a closed curve, we can make use of Stokes theorem to recast the second term in (\ref{eucACT}) as
\begin{equation}
S_{E} = m \int_{\gamma} \sqrt{g_{\mu\nu}\dot{z}^{\mu}\dot{z}^{\nu}} d\tau - qE\mathcal{A}\,,
\end{equation}
where $\mathcal{A}$ denotes the proper area enclosed by the worldline, and $z^\mu{(\tau)} = (\theta{(\tau)},\phi{(\tau)})$. By rotational symmetry, we choose spherical coordinates so that the worldline has constant polar angle $\theta$.
Furthermore, if the loop is traversed $n$ times, we can choose
\begin{equation}
\phi{(\tau)} = 2\pi n \tau\,,\qquad\tau\in(0,1)\,.
\end{equation}
In the Lorentzian signature, $n$ is interpreted as the number of pairs being created.
The total action in this case evaluates to
\begin{equation}\label{Euclideanaction}
S_{E} = \frac{2\pi mn}{H}\sin{\theta} - \frac{2\pi q E n}{H^{2}}(1-\cos{\theta}).
\end{equation}
Extremizing with respect to $\theta$ yields the value
\begin{equation}
\tan{\theta} = \frac{mH}{qE}.
\end{equation}
For this angle, the radius of the circle $y \equiv H^{-1}\sin{\theta}$ is given by
\begin{equation}\label{extremalradius}
y = \frac{m}{\sqrt{q^{2}E^{2}+m^{2}H^{2}}},
\end{equation}
in agreement with the result obtained by \cite{Garriga:1993fh}. It is illuminating to consider the flat space limit, $H\to0$. In this case, we find the radius to be
\begin{equation}
y^{\text{(flat)}} = \frac{m}{qE}.
\end{equation}
This is exactly the distance between the two particles at which their electric potential energy balances their rest mass, allowing the particles to tunnel through the potential barrier to become real. Finally, substituting the extremal value of $y$ back into the action we obtain
\begin{equation}\label{productionratewk}
S_{E} =  \frac{2\pi}{H^{2}} \left(\sqrt{m^{2}H^{2}+q^{2}E^{2}} - qE \right)\,.
\end{equation}

A remark is in order regarding the sign of $q$. When $q >0$, the electric field inside the Wilson loop is smaller than the external field $E$. This corresponds to the screening orientation of the pair, which accelerates away from each other. In the other case ($q<0$), the two charges actually accelerates toward each other, corresponding to the anti-screening orientation. Notice that, in the flat space limit, the action above remains finite for the screening orientation but diverges for the anti-screening orientation. This can be understood as follows: the anti-screening orientation violates conservation of energy, and therefore is forbidden in flat space. In de Sitter space, however, energy needs not be conserved on superhorizon scales, and the anti-screening orientation is allowed. It may seem that for the anti-screening orientation, the two charges will eventually meet each other and annihilate, but this is not the case owing to the expansion of space. Notice also that, due to the compactness of Euclidean dS, which region should be the inside or outside of the circular Wilson loop is ambiguous; in the anti-screening orientation, the outside region is actually smaller than the inside. From here on, we will set $q=1$ and specialize to the screening orientation. In the flat space limit, then, the action becomes
\be
S_{E}^{\text{(flat)}} = \frac{\pi m^{2}n}{E}\,.
\ee
This is exactly the exponent appearing in Schwinger's formula (\ref{Schwingerformula}) for the production rate.

\end{document}